\documentclass[10pt]{iopart}
\usepackage{graphicx}
\usepackage{iopams}
\usepackage{epsfig}
\usepackage{psfrag}

\def\hbarit {{\mathchar'26\mkern-9muh}} 
\begin{document}
\jl{1}
\title{Oscillator model for dissipative QED in an inhomogeneous 
dielectric}
\author{A J van Wonderen and L G Suttorp}
\address{Instituut voor Theoretische Fysica, Universiteit van 
Amsterdam, Valckenierstraat 65, 1018 XE Amsterdam, The Netherlands}

\begin{abstract}

The Ullersma model for the damped harmonic oscillator is coupled to 
the quantised electromagnetic field. All material parameters and 
interaction strengths are allowed to depend on position. The 
ensuing Hamiltonian is expressed in terms of canonical fields, and 
diagonalised by performing a normal-mode expansion. The commutation 
relations of the diagonalising operators are in agreement with the 
canonical commutation relations. For the proof we replace all sums 
of normal modes by complex integrals with the help of the residue 
theorem. The same technique helps us to explicitly calculate the 
quantum evolution of all canonical and electromagnetic fields. We 
identify the dielectric constant and the Green function of the wave 
equation for the electric field. Both functions are meromorphic in 
the complex frequency plane. The solution of the extended 
Ullersma model is in keeping with well-known phenomenological rules 
for setting up quantum electrodynamics in an absorptive and 
spatially inhomogeneous dielectric. To establish this fundamental 
justification, we subject the reservoir of independent harmonic 
oscillators to a continuum limit. The resonant frequencies of the 
reservoir are smeared out over the real axis. Consequently, the 
poles of both the dielectric constant and the Green function unite 
to form a branch cut. Performing an analytic continuation beyond 
this branch cut, we find that the long-time behaviour of the 
quantised electric field is completely determined by the sources of 
the reservoir. Through a Riemann-Lebesgue argument we demonstrate 
that the field itself tends to zero, whereas its quantum 
fluctuations stay alive. We argue that the last feature may have 
important consequences for application of entanglement and related 
processes in quantum devices.     
    
\end{abstract}   

\pacs{42.50.Nn, 03.70.+k, 05.30.-d} 

\submitted 

\maketitle

\section{Introduction}

More than a decade ago, Huttner and Barnett \cite{HB:1992} published a 
valuable contribution to the complicated and long-established 
\cite{JAU:1948} subject of quantum electrodynamics in nonrelativistic 
macroscopic matter. Opting for a canonical setting, they carried out a 
fundamental derivation of the quantised electromagnetic field in an 
absorptive dielectric medium. To open up the possibility of working 
analytically, they avoided making direct contact with the atomic level. 
Instead, they described the properties of the dielectric with the help 
of harmonic oscillators. The influence of absorption was mimicked by 
coupling the dielectric to a continuum of harmonic oscillators. These 
reasonable simplifications gave rise to a so-called damped-polariton 
model that could be exactly solved. The quadratic Hamiltonian was 
diagonalised by means of Fano's method \cite{FAN:1961} and Fourier 
transformation.   

The solution for the quantised electromagnetic field in an absorptive 
dielectric was welcomed by a large community. Phenomenological 
quantisation schemes, the practical value of which was beyond dispute, 
could be put on a solid microscopic foundation. This possibility marked 
the beginning of a period of progress, in which our knowledge of QED in 
or near macroscopic matter was considerably broadened. For a lot of 
different optical media and experimental geometries, ranging from 
magnetic materials to beam splitters, quantisation of the 
electromagnetic field was successfully carried out. At present, the 
macroscopic formulation of QED is well established. A few years ago, it 
was comprehensively reviewed \cite{KNO:2001,LUK:2002}.  

Over the years, it was pointed out several times \cite{YG:1996,DKW:2002} 
that the justification of Huttner and Barnett has one major shortcoming. 
The dielectric is assumed to be homogeneous in space, whereas in many 
experimental situations the electromagnetic field is substantially 
influenced by spatial inhomogeneities. One thus would like to extend the 
solution of the homogeneous damped-polariton model to the case in which 
all model parameters depend on position. This is a far from trivial 
enterprise, because the lack of translational invariance deprives us of 
the possibility to perform Fourier transformation. On the other hand, the 
quadratic character of the Hamiltonian remains intact, so there are no 
mathematical indications for a failure of Fano's procedure when passing 
over to the inhomogeneous case. Indeed, in two companion papers it is 
demonstrated that the inhomogeneous damped-polariton model can be solved 
as well. One can employ either Fano's procedure \cite{SWO:2004}, or an 
alternative method that is based on Laplace transformation \cite{SW:2004}.     

The extension of the work of Huttner and Barnett to dielectrics with 
spatial inhomogeneities does not complete the program of underpinning   
phenomenological quantisation rules. This judgement, which is the main 
motivation for undertaking the present study, is supported by several 
arguments. First, standard works on conservative QED 
\cite{HEI:1954}--\cite{LOU:1973} and classical fields 
\cite{MF:1953,GOL:1980} insist on reducing the collection of canonical 
variables to an enumerable lot before commencing a canonical quantisation 
procedure. In doing so for the inhomogeneous case, we are automatically 
led to a complete set of normal-mode functions, on the basis of which the 
electromagnetic fields can be expanded. We thus identify the functions 
that replace the plane waves of the homogeneous case. Moreover, we find 
the natural decomposition of the Green function \cite{SWO:2004,SW:2004} 
belonging to the wave equation for the electric field. 

Our second reservation with regard to the damped-polariton model concerns 
the loss mechanism. Instead of employing a continuum of oscillators right 
from the beginning, we should postpone the transition to the continuum as 
long as possible. This allows for a scrutiny of the mathematical origin of 
the absorptive behaviour. Also, we can completely clarify the relation 
between irreversibility of the dynamics and causality of the dielectric 
constant. Our third argument pertains to a technical observation on Fano's 
method in the presence of continua. In solving for the diagonalising 
operators one must introduce a formal contribution that is proportional to 
a Dirac delta function. We wish to improve upon this approach. To that 
end, we must refrain from employing distributions \cite{FAN:1961}. Fourth 
and last, we remark that already back in the 1960s quantum dissipation was 
studied with the help of harmonic oscillators  
\cite{SEN:1960}--\cite{DEK:1981}. It is important to 
find out whether the older oscillator models corroborate the predictions 
of the damped-polariton model.         

We can meet the four suggestions advanced above by exchanging the 
continuum of the damped-polariton model for a reservoir containing a 
finite collection of independent harmonic oscillators. In essence, we add 
an electromagnetic sector to the Ullersma model \cite{ULL:1966} for a 
damped harmonic oscillator. As expected, we can still diagonalise the 
Hamiltonian of the extended Ullersma model. Eventually, we come up with a 
mathematical limit that restores the continuum. 

In section 2 we specify the Hamiltonian of our model. In preparation of a 
swift canonical quantisation procedure, we solve the classical evolution 
equations for the canonical fields by invoking a normal-mode expansion. 
Quantisation of the dynamics happens in section 3. With each normal mode 
we associate a quantised harmonic oscillator. Next, making use of complex 
integration, we determine the evolution in time of all canonical fields.   
This permits us to identify the dielectric constant as well as the Green 
function. Subsequently, we turn on absorption by subjecting the reservoir 
to a continuum limit. We witness how poles unite and radically change the 
analytic structure of both the dielectric constant and the Green function. 
We derive the inhomogeneous counterpart of the solution that was obtained 
by Huttner and Barnett. Much attention is devoted to the long-time 
behaviour of the electric field. Section 4 contains a summary and guides 
the reader to the main results of our paper. 

We close this introduction with some technical remarks. In performing 
partial integrations for canonical fields and related quantities, we 
tacitly assume that there are no contributions from the boundaries of 
the dielectric. Our aim to keep the treatment free from distributions 
forces us to utilise a considerable amount of basic function theory; all 
of the accompanying calculus is transferred to two appendices. In an 
attempt to present clear formulas, we omit spatial arguments whenever 
possible. The subscript $L(T)$ denotes that one should take the 
longitudinal (transverse) part of a vector field or tensor field. Last, 
we make use of rationalised mks units throughout.    

\section{Classical treatment}

In this section we derive the classical Hamiltonian of our model. We let us 
be guided by the Lagrange formalism, so our treatment bears a canonical 
character from the very outset. As usual, Hamilton's equations furnish the 
evolution laws for the canonical fields. In solving these, we rely on a  
normal-mode expansion. It appears that the dynamics is governed by an 
eigenvalue problem for a self-adjoint differential operator in three 
dimensions. For a spatially homogeneous dielectric the eigenvectors can be 
constructed from plane waves. To verify that the normal modes neatly 
decouple, we compute the Hamiltonian on the basis of the solutions for the 
canonical fields. 

\subsection{The model} 

The Ullersma model \cite{ULL:1966} describes the interaction between a single 
harmonic oscillator and a reservoir made up by an array of $N$ independent 
harmonic oscillators. To fulfil our purposes, we assign to all $N+1$ 
oscillators a spatial dependence. The corresponding position vector ${\bf r}$  
covers a finite volume $V$. As a further extension of the Ullersma 
configuration we introduce an electromagnetic sector, denoting the electric 
field and magnetic field at time $t$ as ${\bf E}({\bf r},t)$ and 
${\bf B}({\bf r},t)$, respectively. In vacuum these fields give rise to the 
standard electromagnetic Lagrangean density  
\begin{equation} 
{\cal L}_{EM} = \frac{1}{2} \epsilon_0 {\bf E}^2 - \frac{1}{2\mu_0}{\bf B}^2. 
\label{1}
\end{equation} 
The presence of a dielectric medium of volume $V$ is taken into account by 
the privileged harmonic oscillator of the extended Ullersma model. It has 
mass density $\rho ({\bf r})$ and frequency $\omega_0 ({\bf r})$. The field 
${\bf Q}_0({\bf r},t)$ measures its displacement. The Lagrangean density of 
the dielectric reads 
\begin{equation} 
{\cal L}_{D} = \frac{1}{2}\rho \dot{{\bf Q}}_0^2 -\frac{1}{2} \rho \omega_0^2 
{\bf Q}_0^2. \label{2}
\end{equation}
After suitable scaling, all harmonic oscillators of the reservoir have a mass 
density of $\rho ({\bf r})$ as well. Their frequencies and displacement 
fields are equal to $\omega_n ({\bf r})$ and ${\bf Q}_n({\bf r},t)$, 
respectively. From (\ref{2}) we see that 
\begin{equation} 
{\cal L}_{R} = \sum_{n=1}^N \left( \frac{1}{2}\rho \dot{{\bf Q}}_n^2 -
\frac{1}{2} \rho \omega_n^2 {\bf Q}_n^2\right) \label{3}
\end{equation}
is the Lagrangean density of the free reservoir. 

In the electric-dipole approximation the electric field induces in the 
dielectric a polarisation density ${\cal { P}}= -\alpha {\bf Q}_0$, where 
$\alpha$ is positive. In a strictly microscopic theory $- \alpha ({\bf r})$ 
would be a local electronic charge density. The interaction between the 
electromagnetic fields and the dielectric yields a contribution of $-\sigma_s  
\phi + {\bf j}_s\cdot {\bf A}$ to the Lagrangean density. In absence of free 
charges and currents, one may substitute $\sigma_s =-\nabla\cdot {\cal { P}}$ 
and ${\bf j}_s= \dot{{\cal { P}}}$ for the sources. The electromagnetic 
potentials are determined by the definitions ${\bf E} = -\nabla \phi - 
\dot{{\bf A}}$ and ${\bf B} = \nabla \times {\bf A}$, supplemented with the 
choice ${\bf A}_L= 0$, which is equivalent to the Coulomb gauge. 

As long as it allows for energy exchange, the precise form of the interaction 
between the dielectric and the reservoir is not of physical interest to us. 
Introducing a coupling $\beta_n ({\bf r})$, we link the displacement field 
${\bf Q}_0$ to the time derivative $\dot{{\bf Q}}_n$ rather than the field 
${\bf Q}_n$ itself. This departure from the Ullersma model is common in 
studies of dissipative QED \cite{HB:1992}. The two interactions of our model 
make the following contribution to the Lagrangean density: 
\begin{equation} 
{\cal L}_{I} = \alpha (\nabla \phi)\cdot {\bf Q}_0 
-\alpha {\bf A}\cdot \dot{{\bf Q}}_0 
-\sum_{n=1}^N \beta_n {\bf Q}_0\cdot \dot{{\bf Q}}_n. \label{4} 
\end{equation} 
We have added the total derivative $\nabla \cdot (\alpha \phi {\bf Q}_0)$, 
so that the nabla operator acts on the scalar potential. 

As the Lagrangean $L = \int d{\bf r}( {\cal L}_{EM}+{\cal L}_D+{\cal L}_R 
+{\cal L}_I)$ does not depend on $\dot{\phi}$, one immediately finds 
$\nabla \phi = -(\alpha {\bf Q}_0)_L/\epsilon_0$ from the corresponding 
Euler-Lagrange equation. Owing to the Coulomb gauge, the character of the 
first term of (\ref{4}) is purely electrostatic. The remaining Euler-Lagrange 
equations provide us with the inhomogeneous Maxwell equation for the vector 
potential, and the equations of motion for all $N+1$ displacement fields.     

Defining canonical momenta as 
\begin{equation} 
{\bf \Pi} = \frac{\delta L}{\delta \dot{{\bf A}}} 
\hspace{1cm} 
{\bf P}_n = \frac{\delta L}{\delta \dot{{\bf Q}}_n} \label{5} 
\end{equation} 
for $0\le n\le N$, we obtain the Hamiltonian  
\begin{equation}
 H = \int d{\bf r} \left(\dot{{\bf A}}\cdot {\bf \Pi} + \sum_{n=0}^N 
\dot{{\bf Q}}_n\cdot {\bf P}_n \right) -L \label{6}
\end{equation}
as 
\[
\hspace{-14mm}
 H = \int d{\bf r} \left[ \frac{1}{2\epsilon_0}{\bf \Pi}^2 + \frac{1}{2\mu_0}
(\nabla \times {\bf A})^2 + \frac{\alpha^2}{2\rho}{\bf A}^2 
+\frac{1}{2\epsilon_0}(\alpha {\bf Q}_0)_L^2 
+\frac{1}{2\rho} {\bf P}_0^2  
+ \frac{1}{2}\rho \tilde{\omega}_0^2 {\bf Q}_0^2 \rule{0mm}{6mm}\right.
\]
\begin{equation}
\hspace{-14mm}
\left.
+\sum_{n=1}^N \left( \frac{1}{2\rho} {\bf P}_n^2 
+ \frac{1}{2}\rho \omega_n^2 {\bf Q}_n^2 \right) 
+ \frac{\alpha}{\rho} {\bf A}\cdot {\bf P}_0 
+ \sum_{n=1}^N \frac{\beta_n}{\rho} {\bf Q}_0\cdot {\bf P}_n
\right].
\label{7}
\end{equation}   
For the sum $\omega_0^2 + \sum_{n=1}^N \beta_n^2/\rho^2$ the 
abbreviation $\tilde{\omega}_0^2$ will be in use. The Hamiltonian of 
our model being available, we can start investigating the evolution 
of the canonical fields. 

\subsection{Solution of Hamilton's equations} 

Since we work in a finite volume, we may try to unravel the dynamics 
with the help of an enumerable set of independent normal modes. We 
shall see that each mode has a specific spatial structure and 
oscillates at a specific frequency. For that reason, the modes must 
be labelled by both a spatial index $k$ and an  integer $l$, which 
enumerates the mode frequencies $\Omega (k,l)$ for $k$ fixed. The 
index $j$ identifies the canonical fields with the auxiliary 
fields ${\bf Z}_j$ in the following manner:  
\begin{equation} 
\{{\bf \Pi}, {\bf A}, {\bf P}_0, {\bf Q}_0, {\bf P}_n, {\bf Q}_n\} = 
\{{\bf Z}_{1}, {\bf Z}_{2}, {\bf Z}_{3}, {\bf Z}_{4}, 
{\bf Z}_{5n}, {\bf Z}_{6n}\} \label{8} 
\end{equation}
with $n= 1,2,3,\ldots,N$. Now we can put forward our normal-mode 
expansion  
\begin{equation}
{\bf Z}_j({\bf r},t) = \sum_{k,l} c(k,l) {\bf a}_j(k,l;{\bf r}) 
e^{-i\Omega (k,l)t} + \mbox{cc}. \label{9} 
\end{equation}
The frequencies $\Omega (k,l)$ are positive by definition. The 
coefficients $c(k,l)$ could be absorbed in the mode amplitudes 
${\bf a}_j$, but this is inconvenient in view of the quantisation 
procedure lying ahead of us. Because of the Coulomb gauge the 
amplitudes ${\bf a}_{1}$ and ${\bf a}_{2}$ are transverse vector 
fields. 

Upon substituting (\ref{9}) into Hamilton's equations and carefully 
evaluating functional derivatives, one arrives at 
\begin{eqnarray} 
i \Omega {\bf a}_{1} & = & - \frac{1}{\mu_0} \triangle {\bf a}_{2} 
+ \left( \frac{\alpha^2}{\rho} {\bf a}_{2}\right)_T 
+ \left( \frac{\alpha}{\rho} {\bf a}_{3}\right)_T \nonumber \\
i \Omega {\bf a}_{2} & = & - \frac{1}{\epsilon_0} 
{\bf a}_{1} \nonumber \\
i \Omega {\bf a}_{3} & = & \rho \tilde{\omega}_0^2 {\bf a}_{4} 
+ \frac{\alpha}{\epsilon_0} \left(\alpha {\bf a}_{4}\right)_L 
+\sum_{m=1}^N \frac{\beta_m}{\rho} {\bf a}_{5m} \nonumber \\
i \Omega {\bf a}_{4} & = & -\frac{\alpha}{\rho} {\bf a}_{2} 
-\frac{1}{\rho}{\bf a}_{3} \nonumber \\
i \Omega {\bf a}_{5n} & = & \rho \omega_n^2 {\bf a}_{6n} \nonumber \\
i \Omega {\bf a}_{6n} & = & -\frac{\beta_n}{\rho} {\bf a}_{4} 
-\frac{1}{\rho} {\bf a}_{5n}  \label{10}
\end{eqnarray} 
with $1\le n\le N$. The arguments of ${\bf a}_j$ and $\Omega$ are 
identical to those appearing in (\ref{9}). We emphasize that the model 
parameters $\alpha$, $\beta_n$, $\rho$, $\tilde{\omega}_0$, and 
$\omega_n$ depend on position. 

We set out to derive a wave equation for the amplitude 
${\bf e}(k,l;{\bf r})$ of the electric field. The expansion 
\begin{equation} 
{\bf E}({\bf r},t) = \sum_{k,l} 
c(k,l) {\bf e}(k,l;{\bf r}) e^{-i\Omega (k,l)t} 
+ \mbox{cc} \label{12}
\end{equation} 
should match with the definition of ${\bf E}$ in terms of the 
electromagnetic potentials. This brings us to the prescription  
\begin{equation}
\epsilon_0 {\bf e} = -{\bf a}_{1} + \left(\alpha {\bf a}_{4}\right)_L. 
\label{11}
\end{equation} 
The above relation enables us to present the solution of the algebraic 
part of (\ref{10}) as  
\begin{eqnarray} 
{\bf a}_{1} = -\epsilon_0 {\bf e}_T  & \hspace{1cm} &
{\bf a}_{2} = -\frac{i}{\Omega }{\bf e}_T \nonumber \\ 
{\bf a}_{3} = \frac{i\alpha}{\Omega}{\bf e}_T 
-\frac{i\alpha\Omega}{h(\Omega)} {\bf e} & \hspace{1cm} & 
{\bf a}_{4} = \frac{\alpha}{\rho h(\Omega)}{\bf e} \nonumber \\
{\bf a}_{5n} = 
\frac{\alpha \beta_n\omega_n^2}{\rho h(\Omega)(\Omega^2-\omega_n^2)} 
{\bf e}  & \hspace{1cm} & 
{\bf a}_{6n} = 
\frac{i\alpha \beta_n\Omega}{\rho^2 h(\Omega)(\Omega^2-\omega_n^2)} 
{\bf e}  . \label{13}
\end{eqnarray} 
The new function $h(s)$ implicitly depends on position, and is given by 
\begin{equation}
h(s) = s^2 - \tilde{\omega}_0^2 + \sum_{n=1}^N 
\frac{\beta_n^2 \omega_n^2}{\rho^2 (\omega_n^2 -s^2)}. \label{14}
\end{equation} 
In the next section, the real-valued argument $s$ will be replaced by a 
complex variable. A swift verification of (\ref{13}) may take place 
through substitution into (\ref{10}). Indeed all five algebraic equations 
are satisfied if (\ref{11}) is used. 

The first equation of (\ref{10}), containing the Laplacian operator, 
has not been considered as yet. After substitution of (\ref{13}) and 
employment of (\ref{11}) it reduces to 
\begin{equation} 
c^2 \nabla \times \left(\nabla \times {\bf e}\right) = 
\Omega^2 \varepsilon (\Omega) {\bf e}. \label{14a} 
\end{equation}   
This is the standard wave equation for the electric field, which appears 
in electromagnetic theory. The function   
\begin{equation}
\varepsilon (s) = 1-\frac{\alpha^2}{\epsilon_0 \rho h(s)} \label{19} 
\end{equation}
plays the role of dielectric constant. Its spatial dependence is made 
explicit in the following.  

In (\ref{14a}) a linear differential operator is at work. It is given  
by    
\begin{equation} 
L(s;{\bf r}) {\bf v}({\bf r}) = 
-c^2\nabla \times \left[\nabla \times {\bf v}({\bf r})\right] 
+ s^2 \varepsilon (s;{\bf r}) {\bf v}({\bf r}). \label{18}
\end{equation} 
The parameter $s$ is arbitrary, but still real-valued. The 
vector field ${\bf v}$ belongs to a Hilbert space of complex and 
square integrable functions, with scalar product defined as    
\begin{equation} 
\langle {\bf v}_1, {\bf v}_2\rangle = \int d {\bf r}\,\, 
{\bf v}_1^{\ast}({\bf r})\cdot {\bf v}_2({\bf r}). \label{18a} 
\end{equation} 
The integration is confined to the finite volume $V$. Since $L(s)$ 
is self-adjoint, its eigenvalues $\lambda (k,s)$ are real. Second, its 
eigenvectors ${\bf u}(k,s)$ possess the orthonormality property  
\begin{equation} 
\langle {\bf u}(k,s), {\bf u}(k',s)\rangle = \delta_{kk'}. \label{21} 
\end{equation}
Obviously, (\ref{21}) is not affected by degeneracy. Eigenspaces of 
higher dimension should be subjected to an orthonormalisation 
procedure. Third, the eigenvectors make up a basis for the Hilbert 
space. We have a decomposition of the tensorial delta function at our 
disposal, given by 
\begin{equation} 
\sum_k {\bf u}^{\ast}(k,s;{\bf r}^{\prime}) {\bf u}(k,s;{\bf r}) 
= \mbox{{\boldmath $\delta$}}({\bf r}^{\prime}-{\bf r}) . \label{23} 
\end{equation} 
The reality of the left-hand side follows from the invariance of the 
tensorial delta function under interchange of its arguments and 
indices.  Last, as $L(s)$ is quadratic in the parameter $s$, the 
eigenvectors for $s$ and $-s$ can be chosen identical to each other, 
so that the relation 
\begin{equation} 
{\bf u}(k,-s;{\bf r}) = {\bf u}(k,s;{\bf r}) \label{40} 
\end{equation} 
may be assumed. 
 
Of course, for $s$ equal to a mode frequency $\Omega$, the eigenvalue 
problem  
\begin{equation} 
L(s;{\bf r}) {\bf u}(k,s;{\bf r}) = \lambda (k,s) {\bf u}(k,s;{\bf r})   
\label{16} 
\end{equation} 
should be in keeping with the wave equation (\ref{14a}). We therefore 
make the choice   
\begin{equation} 
{\bf e}(k,l;{\bf r}) =  w(k,l) {\bf u}(k, \Omega(k,l);{\bf r}).  
\label{15} 
\end{equation} 
The weight $w$ will be evaluated in due course. We furthermore require  
that the eigenvalue $\lambda (k,s)$ be equal to zero if $s$ coincides 
with a mode frequency. In other words, by solving the equation  
\begin{equation} 
\lambda (k,s) =0 
\label{17}
\end{equation} 
with $s$ real-valued, we gather all mode frequencies. From (\ref{21}) and 
(\ref{16}) one infers  
\begin{equation}
\lambda (k,s) = \langle {\bf u}(k,s),s^2 \varepsilon (s) 
{\bf u}(k,s)\rangle 
-c^2 \langle \nabla \times {\bf u}(k,s), \nabla \times {\bf u}(k,s)\rangle 
. \label{24}
\end{equation}
A partial integration has been performed. In appendix A we use (\ref{24}) 
to take a closer look at the solutions of (\ref{17}). From (\ref{40}) we 
see that $\lambda (k,s)$ is even in $s$, so only pairs $s= \pm \Omega$ 
occur. The negative solutions generate the cc terms of (\ref{9}). 

In summary, Hamilton's equations have been solved by means of a normal-mode 
expansion. The spatial structure of the normal modes is described by the 
eigenvalue problem (\ref{16}), whereas the mode frequencies are specified 
by the constraint (\ref{17}). Now we should keep our promise to demonstrate   
that the modes do not interact with each other. 
 
\subsection{Computation of the Hamiltonian} 

We insert the normal-mode expansion (\ref{9}) for all canonical fields into 
the Hamiltonian (\ref{7}). Subsequently, we employ (\ref{13}) and (\ref{15}) 
to convert all amplitudes ${\bf a}_j(k,l)$ into eigenvectors 
${\bf u}(k,\Omega)$. The arguments $(k,l)$ of $\Omega$ are omitted. 
Likewise, by $\Omega'$ the mode frequency $\Omega (k',l')$ is meant. If we  
perform a partial integration, the wave equation (\ref{14a}) allows us to 
get rid of all differential operators. The longitudinal contributions of 
(\ref{7}) can be eliminated via the relation 
\begin{equation} 
\left[ \varepsilon (\Omega) {\bf u}(k,\Omega)\right]_L = 0 \label{26}
\end{equation}
which is a consequence of (\ref{14a}) and (\ref{15}).

The Hamiltonian is equal to a sum over indices $k,l,k',l'$. We can 
interchange these to make all summands maximally symmetric. Then two classes 
of summands remain: those oscillating rapidly at frequency $\Omega +\Omega'$, 
and those oscillating slowly at frequency $\Omega -\Omega'$. For each summand 
of the first class there is a complex conjugate, which may be ignored in the  
sequel. For clarity we remark that summands oscillating at frequency 
$-\Omega + \Omega'$ can be transferred to the second class. A simple 
interchange of summation indices suffices.  

The summands of rapid oscillation will be treated first. All of these can  
be expressed in terms of $h$ functions, owing to the relation 
\begin{equation} \hspace{-3mm} 
\sum_{n=1}^N 
\frac{\beta_n^2 \omega_n^2 (\omega_n^2 - s s')}
{\rho^2 (\omega_n^2-s^2)(\omega_n^2-s'^2)} = 
\frac{sh(s)+s'h(s')}{s+s'} -s^2-s'^2+ss'+\tilde{\omega}_0^2. \label{27a}
\end{equation} 
The sum on the left-hand side originates from (\ref{7}). The dummies $s$ and 
$s'$ are set equal to $\Omega$ and $\Omega'$, respectively. Verification of 
(\ref{27a}) is straightforward after substitution  of the definition 
(\ref{14}) on the right-hand side. If we eliminate all $h$ functions in 
favour of dielectric functions (\ref{19}), the summands of rapid oscillation 
yield a form that vanishes after use of the identity 
\begin{equation} 
\langle {\bf u}^{\ast}(k',\Omega'), \Omega'^2 \varepsilon(\Omega') 
{\bf u}(k,\Omega)-\Omega^2 \varepsilon(\Omega){\bf u}(k,\Omega)\rangle =0. 
\label{27}
\end{equation}
One proves (\ref{27}) by taking the scalar product of 
${\bf u}^{\ast}(k',\Omega')$ and the wave equation (\ref{14a}). Once again, 
partial integration is indispensable. 

We have found that only the summands of slow oscillation contribute to the 
Hamiltonian. As long as $\Omega$ differs from $\Omega'$, these summands can 
be handled in the same manner as discussed above. There is only one 
difference. In order to finalise the calculation, one needs the identity 
\begin{equation} 
\langle {\bf u}(k',\Omega'), \Omega'^2 \varepsilon(\Omega') 
{\bf u}(k,\Omega)-\Omega^2 \varepsilon(\Omega){\bf u}(k,\Omega)\rangle =0 
\label{27c}
\end{equation}
instead of (\ref{27}). The proofs of (\ref{27}) and (\ref{27c}) follow the 
same path. Altogether, for $\Omega \ne \Omega'$ the summands of slow 
oscillation add up to zero as well. 

The case $\Omega = \Omega'$ calls for employment of the relation  
\begin{equation}
\sum_{n=1}^N 
\frac{\beta_n^2 \omega_n^2 (\omega_n^2 + s^2)}
{\rho^2 (\omega_n^2-s^2)^2} = 
\frac{d\,sh(s)}{ds} -3s^2+\tilde{\omega}_0^2. \label{27b}
\end{equation} 
Now elimination of $h$ in favour of $\varepsilon$ no longer produces a null 
result. We are led to   
\begin{equation} 
\hspace{-25mm} 
H = 2 \epsilon_0 
\sum_{k,l,k',l'}\rule{0mm}{3mm}^{\!\!\prime} 
c(k,l)c^{\ast}(k',l') w(k,l) w^{\ast}(k',l')  
\left\langle {\bf u}(k',s), 
\frac{ds^2\varepsilon (s)}{ds^2}
{\bf u}(k,s)\right\rangle_{s=\Omega(k,l)} . \label{28}
\end{equation} 
Because of the constraint $\Omega =\Omega'$ the summation carries a 
prime. As the wave equation (\ref{14a}) is only valid for discrete values 
of $\Omega$, it may not be differentiated with respect to this 
variable. Therefore, further manipulations of (\ref{28}) must take place 
in a meticulous manner. From partial integration and (\ref{14a}) we learn  
\begin{equation}
\left\langle {\bf u}(k',s), 
\frac{ds^2\varepsilon (s)}{ds^2}
{\bf u}(k,s)\right\rangle_{s=\Omega} = \left[\frac{d}{ds^2} 
\left\langle {\bf u}(k',s), L(s) 
{\bf u}(k,s)\right\rangle\right]_{s=\Omega} \label{29}
\end{equation} 
with the condition $\Omega = \Omega'$ in force. The eigenvalue equation 
(\ref{16}) holds true for all real values of $s$, so it may be utilised  
on the right-hand side of (\ref{29}). Bearing in mind that the 
eigenvectors are orthonormal, and moreover, that at fixed $k$ the 
correspondence between $\Omega (k,l)$ and $l$ is one-to-one, we can finish 
our computation of the Hamiltonian. The diagonal form 
\begin{equation} 
H = 2 \epsilon_0 \sum_{k,l} |c(k,l)|^2 |w(k,l)|^2 
\left[\frac{d\lambda (k,s)}{ds^2}\right]_{s=\Omega (k,l)}  \label{30} 
\end{equation} 
is found. In accordance with our expectations, the normal modes do not 
interact with each other.  

We choose the weights as 
\begin{equation}
w(k,l) = \epsilon_0^{-1/2} \Omega (k,l) 
\left\{\left[
\frac{d\lambda (k,s)}{ds}\right]_{s=\Omega (k,l)}\right\}^{-1/2} \label{32} 
\end{equation} 
so that (\ref{30}) becomes 
\begin{equation} 
H = \sum_{k,l} \Omega (k,l) |c(k,l)|^2 . \label{31}
\end{equation} 
This is the Hamiltonian of an enumerable collection of independent harmonic 
oscillators. The quantity $c(k,l)$ is sometimes called the amplitude of the 
normal mode $k,l$ \cite{LOU:1973}. With the normal-mode expansion (\ref{9}) 
and the Hamiltonian (\ref{31}) in hand, we fully understand how the 
classical dynamics of our model works.   

\subsection{Homogeneous dielectric} 

Analytic solution of the eigenvalue problem (\ref{16}) will be impossible, 
except for special cases. One of these is a spatially homogeneous dielectric.  
The model parameters no longer depend on position. One can dispose of the 
parameter $s$ figuring in the eigenvectors of (\ref{16}). If one does not 
care for rotational or other spatial symmetries, then a simple solution for 
the eigenvectors is provided by plane waves. 

A cube of side $L$ serves as the volume $V$ of the dielectric. The spatial 
index $k$ splits up into a polarisation index $\mu=1,2,3$ and a wave vector 
${\bf q}= 2\pi {\bf m}/L$, with $m_j$ any integer. The eigenvectors 
${\bf u}(k;{\bf r})$ are given by $L^{-3/2} \hat{{\bf o}}_{\mu}\exp 
(i{\bf q}\cdot {\bf r})$, where $\hat{{\bf o}}_3$ equals $\hat{{\bf q}}$ and 
the set $\{\hat{{\bf o}}_1,\hat{{\bf o}}_2,\hat{{\bf o}}_3\}$ is orthonormal. 
In order to fulfil (\ref{23}) the continuum limit $L\rightarrow\infty$ should 
be taken.   

The solution of (\ref{16}) and (\ref{24}) is composed of a longitudinal part 
and a transverse part. The longitudinal and transverse mode frequencies obey 
the equations 
\begin{equation} 
\varepsilon (\Omega) = 0 \hspace{1cm} \Omega^2 \varepsilon (\Omega) = 
c^2 {\bf q}^2 . \label{25}
\end{equation}
The fact that the dielectric is invariant under rotations causes the 
degeneracy in (\ref{25}).  

\section{Quantum treatment} 

In the previous section we recognised that the classical evolution of our 
extended Ullersma model is controlled by an enumerable set of independent 
harmonic oscillators. For the quantisation of each oscillator we 
resort to the old method of Dirac. The advantage is that the quantised 
Hamiltonian is delivered to us in diagonal shape. Since we work in a 
canonical setting throughout, we must ascertain that the canonical operators 
obey the canonical commutation relations. Completion of this job teaches us 
that discrete sums over normal modes can be transformed into complex 
integrals, if in the eigenvalue problem (\ref{16}) the parameter $s$ is  
replaced by a complex variable. The transition to complex integration 
is the key to disclosing the quantum evolution of the extended Ullersma 
model. For all Heisenberg operators, including the electric field and the 
electric displacement, integral representations are obtained. The dynamics 
is in agreement with Heisenberg's equations. To bring about dissipation in 
the dielectric, we make use of a continuum limit, which allots to the 
reservoir an uncountable number of degrees of freedom. Once the energy sink 
is activated, we can identify the causal dielectric function. 

\subsection{Dirac quantisation} 

From here onwards, all capitals refer to quantummechanical operators, the 
Green function ${\sf G}$, the operator $L$, and the frequency $\Omega$ 
excepted. Following Dirac, we associate with each normal mode $k,l$ of 
energy  $\Omega (k,l) |c(k,l)|^2$ a harmonic oscillator. We define ladder 
operators through 
\begin{equation} 
c(k,l) \rightarrow \hbarit^{1/2} C(k,l) \hspace{1cm} 
c^{\ast}(k,l) \rightarrow \hbarit^{1/2} C^{\dagger}(k,l) \label{33}
\end{equation}
and quantise by postulating 
\begin{equation} 
\left[C(k,l),C^{\dagger}(k',l')\right] = \delta_{kk'}\delta_{ll'} 
\hspace{1cm} 
\left[C(k,l),C(k',l')\right] =0. \label{34}
\end{equation} 
Upon symmetrizing properly, we obtain the quantised counterpart of the 
Hamiltonian (\ref{31}) as  
\begin{equation} 
H = \frac{\hbarit}{2} \sum_{k,l} \Omega (k,l) \left[ C^{\dagger}(k,l)C(k,l) 
+ C(k,l)C^{\dagger}(k,l)\right]. \label{35}
\end{equation} 
Each eigenvalue $\Omega$ is positive. From (\ref{9}) and (\ref{33}) we deduce 
that all canonical operators are represented by the expansion 
\begin{equation}
{\bf Z}_j({\bf r},t) = \hbarit^{1/2}\sum_{k,l} {\bf a}_j(k,l;{\bf r})C(k,l) 
e^{-i\Omega (k,l)t} + \mbox{hc}. \label{36}
\end{equation} 
The amplitudes ${\bf a}_j$ are completely determined by the results 
established in the previous section. In calculating the quantised Hamiltonian, 
one may also depart from (\ref{7}). After substitution of (\ref{36}) into 
(\ref{7}), one enters a rather lengthy road. The manipulations are 
essentially the same as in section 2.3, so there is no need for any comments. 
One indeed retrieves the diagonal form (\ref{35}).  

\subsection{Canonical commutation relations} 

On our way to the swift quantisation procedure (\ref{34}), we let us be 
guided by the Lagrange formalism. Therefore, one may rightfully ask whether 
all is well with the canonical commutation relations. These read 
\begin{eqnarray} 
\left[ {\bf A}({\bf r}',t), {\bf \Pi}({\bf r},t)\right] & = & 
i\hbarit \mbox{{\boldmath $\delta$}}_T({\bf r}'-{\bf r}) \nonumber \\
\left[ {\bf Q}_0({\bf r}',t), {\bf P}_0({\bf r},t)\right] & = &
i\hbarit \mbox{{\boldmath $\delta$}} ({\bf r}'-{\bf r}) \nonumber \\
\left[ {\bf Q}_{n'}({\bf r}',t), {\bf P}_n({\bf r},t)\right] & = & 
i\hbarit \delta_{n'n} \mbox{{\boldmath $\delta$}} ({\bf r}'-{\bf r}). 
\label{37}
\end{eqnarray}
All other commutators of canonical fields equal zero. In the following  
we shall only verify the upper two commutators (\ref{37}). For the 
other 19 cases new problems do not arise. Incidentally, for some  
commutators verification is trivial. 

In virtue of (\ref{34}) and (\ref{36}) the upper two conditions 
(\ref{37}) can be brought onto the form 
\begin{equation} 
\hspace{-20mm}
\sum_{k,l} \frac{\epsilon_0}{\Omega (k,l)} {\bf e}^{\ast}_T(k,l;{\bf r'}) 
{\bf e}_T(k,l;{\bf r}) + \mbox{cc}  =  
\mbox{{\boldmath $\delta$}}_T({\bf r}'-{\bf r})  \label{38} 
\end{equation} 
\begin{equation}
\hspace{-20mm}   
\sum_{k,l} 
\frac{\alpha ({\bf r})\alpha ({\bf r}'){\bf e}^{\ast}(k,l;{\bf r'})} 
{\rho ({\bf r}')\Omega (k,l) h(\Omega (k,l);{\bf r}')} 
\left[ 
\frac{\Omega^2(k,l){\bf e}(k,l;{\bf r})}{h(\Omega (k,l);{\bf r})} 
-  {\bf e}_T(k,l;{\bf r})\right] + \mbox{cc}  =  
\mbox{{\boldmath $\delta$}}({\bf r}'-{\bf r}) . 
\label{38a} 
\end{equation} 
The relations (\ref{15}) and (\ref{32}) suggest a recourse to the residue 
theorem. In moving to the complex plane we observe the rule 
\begin{equation}
{\bf u}(k,s;{\bf r})\rightarrow {\bf u}(k,z;{\bf r}) \hspace{1cm} 
{\bf u}^{\ast}(k,s;{\bf r}) \rightarrow 
[{\bf u}(k,z^{\ast};{\bf r})]^{\ast} 
\label{39a} 
\end{equation} 
where ${\bf u}(k,z;{\bf r})$ is the solution of (\ref{40}) and (\ref{16}) 
with the replacement $s\rightarrow z$ carried out. In appendix A we argue 
that in (\ref{39a}) two entire functions of $z$ figure. Moreover, we make 
plausible that the function $\lambda^{-1}(k,z)$ is meromorphic, and 
that its poles are given by $z=\pm \Omega (k,l)$, for $k$ fixed.   

Upon performing in (\ref{38}) and (\ref{38a}) the transition to complex 
integration, we find the following three sufficient conditions:  
\begin{equation} 
\hspace{-12mm} 
\sum_k \int_{C_1} \frac{dz}{\pi i} \frac{z}{\lambda (k,z)} 
[{\bf u}_T(k,z^{\ast};{\bf r}')]^{\ast} 
 {\bf u}_T(k,z;{\bf r})  = 
\mbox{{\boldmath $\delta$}}_T({\bf r}'-{\bf r})  
\label{39}
\end{equation}
\begin{equation} 
\hspace{-12mm}
\sum_k \int_{C_1} dz \frac{z}{h(z;{\bf r}')\lambda (k,z)} 
[{\bf u}(k,z^{\ast};{\bf r}')]^{\ast} {\bf u}_T(k,z;{\bf r})  =  0  
\label{42c}  
\end{equation} 
\begin{equation} 
\hspace{-12mm}
\sum_k \int_{C_1} \frac{dz}{\pi i} 
\frac{z^3}{h(z;{\bf r})h(z;{\bf r}')\lambda (k,z)} 
[{\bf u}(k,z^{\ast};{\bf r}')]^{\ast} 
{\bf u}(k,z;{\bf r}) = 
\frac{\epsilon_0 \rho({\bf r})}{\alpha^2({\bf r})}
\mbox{{\boldmath $\delta$}}({\bf r}'-{\bf r}). 
\label{42}
\end{equation}
The contour $C_1$ is composed of a set of circles running in
counterclockwise sense (see figure B1 in appendix B). The $l$th circle
encloses the pole on the positive real axis at $z=\Omega (k,l)$, with $k$
fixed. Everywhere else in the interior of $C_1$ the integrands are
analytic, because the circles can be chosen as small as we like. Hence, use
of the residue theorem in (\ref{39})--(\ref{42}) indeed reproduces the sums
over mode frequencies that are contained in (\ref{38}) and (\ref{38a}).

The conditions (\ref{39})--(\ref{42}) can be proved by performing a 
series of contour deformations. We defer this purely technical exercise to 
appendix B. Right now the reader should stay focused on the 
transition to complex integration, as practised above. We plan to exploit 
that skill in computing the time evolution of the canonical fields and 
other operators of physical interest. This is the goal of the next 
subsection. 

\subsection{Solution of the extended Ullersma model} 

To get a full picture of the dynamics of our quantum system, we have to 
specify the evolution of any set of initial canonical operators. In short, 
we have to establish the mapping between the times $t=0$ and $t$ for all 
canonical operators. To that end, we observe that any quantum operator can be 
written as a linear combination of the canonical operators at time zero. We 
apply this statement to $C^{\dagger}(k,l)$, and use (\ref{36}) as well as 
(\ref{37}) to make all coefficients explicit. We are led to the expansion 
\[
\hspace{-25mm}
C^{\dagger}(k,l) = \frac{i}{\hbarit^{1/2}} \int d{\bf r}\left[\rule{0mm}{6mm}
{\bf a}_{1}(k,l;{\bf r})\cdot {\bf A}({\bf r},0)  
-{\bf a}_{2}(k,l;{\bf r})\cdot {\bf \Pi}({\bf r},0)
+{\bf a}_{3}(k,l;{\bf r})\cdot {\bf Q}_0({\bf r},0) \right. 
\]
\begin{equation} 
\hspace{-25mm}
\left.
-{\bf a}_{4}(k,l;{\bf r})\cdot {\bf P}_0({\bf r},0)
+\sum_{n=1}^N {\bf a}_{5n}(k,l;{\bf r})\cdot {\bf Q}_n({\bf r},0) 
-\sum_{n=1}^N {\bf a}_{6n}(k,l;{\bf r})\cdot {\bf P}_n({\bf r},0)\right]. 
\label{43}
\end{equation}
One may wonder whether the representation (\ref{43}) complies with the 
quantisation prescription (\ref{34}). By invoking (\ref{13}), (\ref{15}), 
and (\ref{37}), one derives a set of consistency relations that is 
equivalent to (\ref{27}), (\ref{29}), and (\ref{32}). 

After substitution of (\ref{43}) into (\ref{36}) and use of the 
normalisation (\ref{32}), we can express the vector potential as 
\begin{equation} 
\hspace{-20mm}
{\bf A}({\bf r},t) = \frac{ic^2}{\epsilon_0} \sum_{k,l} \int d{\bf r}' 
\left\{ e^{-ist} 
\frac{{\bf u}_T(k,s;{\bf r}) {\bf u}^{\ast}(k,s;{\bf r}')\cdot 
{\bf j}(s;{\bf r}')}
{d\lambda (k,s)/ds}
\right\}_{s=\Omega (k,l)} +\mbox{hc}. \label{44a} 
\end{equation} 
The source vector must be constructed from the initial canonical fields. 
One has 
\[
\hspace{-25mm}
c^2{\bf j}(s;{\bf r}) = -i\epsilon_0 s {\bf A}({\bf r},0) +
{\bf \Pi}({\bf r},0) 
+\left[\alpha ({\bf r}){\bf Q}_0({\bf r},0)\right]_T 
-\frac{s^2 \alpha({\bf r}){\bf Q}_0({\bf r},0)}{ h(s;{\bf r})} 
 \]
\begin{equation} 
\hspace{-25mm} 
-\frac{is\alpha ({\bf r}){\bf P}_0({\bf r},0)}
{\rho ({\bf r}) h(s;{\bf r})}
+\sum_{n=1}^N 
\frac{s\alpha ({\bf r})\beta_n ({\bf r})}
{\rho ({\bf r})h(s;{\bf r})[s^2-\omega_n^2 ({\bf r})]} 
\left[i\omega_n^2({\bf r}){\bf Q}_n({\bf r},0) -
\frac{s}{\rho ({\bf r})} {\bf P}_n({\bf r},0)\right]. \label{45}
\end{equation} 
The intermediate result (\ref{44a}) paves the way for the residue 
theorem, in a similar vein as before. The following result is 
reached: 
\begin{equation} 
\hspace{-5mm}
{\bf A}({\bf r},t) =  \int d{\bf r}' \int_{C_3} 
\frac{dz}{2\pi \epsilon_0} e^{-izt} {\sf G}_T(z;{\bf r},{\bf r}')\cdot 
{\bf j} (z;{\bf r}'). \label{44}
\end{equation}
The operation $T$ refers to the argument ${\bf r}$. 
The contour $C_3$ encloses the real axis by means of two straight lines 
running from $+\infty +i\eta $ to $-\infty +i\eta$, and from $-\infty 
-i\eta $ to $+\infty -i\eta$, where $\eta$ is infinitesimally positive.
From (\ref{16}) we see that the Green function, defined as  
\begin{equation} 
{\sf G}(z;{\bf r},{\bf r}') = \sum_k \frac{c^2}{\lambda (k,z)}  
{\bf u}(k,z;{\bf r})\left[{\bf u}(k,z^{\ast};{\bf r}')\right]^{\ast} 
\label{44b} 
\end{equation} 
satisfies the partial differential equation 
\begin{equation} 
c^{-2}L(z;{\bf r}) {\sf G}(z;{\bf r},{\bf r}') = 
\mbox{{\boldmath $\delta$}} ({\bf r}-{\bf r}') \label{44c} 
\end{equation} 
where $z$ must lie on the contour $C_3$. As shown in appendix B, the 
integrand of (\ref{44}) has the same analytic structure as 
$\lambda^{-1}(k,z)$. From (\ref{14}) it is clear that in 
(\ref{45}) the factor of $[z^2-\omega_n^2({\bf r})]^{-1}$ does not give 
rise to any poles. Hence, in (\ref{44}) the contour $C_1$ could be 
exchanged for $C_3$ without paying a price. 

For the displacement field ${\bf Q}_0$ the discrete solution is given by 
\begin{equation} 
\hspace{-25mm}
{\bf Q}_0({\bf r},t) = - \sum_{k,l} \int d{\bf r}' 
\left\{ \frac{c^2 s\alpha ({\bf r}) e^{-ist}}{\epsilon_0 \rho ({\bf r}) 
h(s;{\bf r})}  
\frac{{\bf u}(k,s;{\bf r}) {\bf u}^{\ast}(k,s;{\bf r}')\cdot 
{\bf j}(s;{\bf r}')}
{d\lambda (k,s)/ds}
\right\}_{s=\Omega (k,l)} +\mbox{hc}. \label{46b} 
\end{equation} 
The transition to complex integration is immediate, provided that $C_1$ is 
chosen as contour. In deforming the latter to $C_3$, attention must be paid 
to denominators in which a double factor of $h$ figures. Correction terms show 
up, which can be found by eliminating the nasty denominators with the help of 
the auxiliary function ${\sf f}$, defined in appendix B. One should proceed
along the same lines as for the proof of (\ref{42}). This results in  
\[
\hspace{-15mm}
{\bf Q}_0({\bf r},t) = - \int d{\bf r}' \int_{C_3} 
\frac{dz}{2\pi i \epsilon_0} 
\frac{z\alpha ({\bf r})e^{-izt}}{\rho ({\bf r}) h(z;{\bf r})} 
{\sf G}(z;{\bf r},{\bf r}')\cdot {\bf j} 
(z;{\bf r}') 
\]
\begin{equation}
\hspace{-15mm}
-\int_{C_3}\frac{dz}{2\pi i} 
\frac{e^{-izt}}{z\alpha ({\bf r})} \left\{c^2 {\bf j}(z;{\bf r})+i\epsilon_0 z 
{\bf A}({\bf r},0) -{\bf \Pi}({\bf r},0)-\left[\alpha ({\bf r}) 
{\bf Q}_0({\bf r},0)\right]_T\right\}.
\label{46}
\end{equation}
The local character of the correction term stems from application of the 
completeness relation (\ref{23}).  

The displacement field of the reservoir contains two local correction 
terms. The solution comes out as     
\[
\hspace{-25mm}
{\bf Q}_n({\bf r},t) =  -\int d{\bf r}' \int_{C_3} 
\frac{dz}{2\pi \epsilon_0} 
\frac{z^2\alpha ({\bf r})\beta_n ({\bf r}) e^{-izt}}
{\rho^2 ({\bf r}) h(z;{\bf r})[z^2-\omega_n^2({\bf r})]} 
{\sf G}(z;{\bf r},{\bf r}')\cdot {\bf j} (z;{\bf r}') 
\]
\[
\hspace{-25mm}
-\int_{C_3}\frac{dz}{2\pi } 
\frac{\beta_n({\bf r}) e^{-izt}}
{\alpha ({\bf r})\rho({\bf r})[z^2-\omega_n^2({\bf r})]} 
\left\{c^2 {\bf j}(z;{\bf r})+i\epsilon_0 z 
{\bf A}({\bf r},0) -{\bf \Pi}({\bf r},0)-\left[\alpha ({\bf r}) 
{\bf Q}_0({\bf r},0)\right]_T\right\}
\]
\begin{equation}
\hspace{-25mm}
+\int_{C_3}\frac{dz}{2\pi i} 
\frac{e^{-izt}}{[z^2-\omega_n^2 ({\bf r})]} 
\left[ z{\bf Q}_n({\bf r},0) +\frac{i}{\rho ({\bf r})} 
{\bf P}_n({\bf r},0)\right]  
\label{46a}
\end{equation} 
where one has $n=1,2,3,\ldots,N$. The last correction term covers 
the special case of $\alpha = \beta_n =0$. The fields ${\bf A}$, 
${\bf Q}_0$, and ${\bf Q}_n$ provide us with the following 
solutions for the canonical momenta: 
\begin{equation} 
{\bf \Pi}= \epsilon_0 \dot{{\bf A}} \hspace{10mm} 
{\bf P}_0 = \rho \dot{{\bf Q}}_0 -\alpha {\bf A} \hspace{10mm} 
{\bf P}_n = \rho \dot{{\bf Q}}_n -\beta_n {\bf Q}_0 \label{46f} 
\end{equation} 
with $n=1,2,3,\ldots,N$. These expressions are obtained from 
the definitions (\ref{5}).  

All solutions for the canonical operators reproduce the initial 
condition if the choice $t=0$ is made. This can be verified on the 
basis of the identities underlying the canonical commutation 
relations, such as (\ref{39}) and (\ref{42}). The transformation 
$z\rightarrow -z^{\ast}$ shows that the condition of 
self-adjointness is respected as well. One needs the symmetry 
property  
\begin{equation} 
\left[{\sf G}(-z^{\ast};{\bf r},{\bf r}')\right]^{\ast} = 
{\sf G}(z;{\bf r},{\bf r}') \label{46c} 
\end{equation} 
where $z$ belongs to $C_3$. For the proof one combines (\ref{44b}) 
with the reciprocity relation 
\begin{equation} 
{\sf G}(z;{\bf r}',{\bf r})_{ji} = {\sf G}(z;{\bf r},{\bf r}')_{ij} 
. \label{46d} 
\end{equation} 
This last result follows by taking the scalar product of 
(\ref{44c}) with ${\sf G}(z;{\bf r},{\bf r}'')$, and 
performing a partial integration. 

All canonical fields obey the Heisenberg equation 
$i\hbarit \dot{{\bf Z}}_j = [{\bf Z}_j,H]$. To demonstrate this, 
one first calculates the commutator with $H$ on the basis of the 
canonical commutation relations. Subsequently, one employs 
(\ref{46f}) to eliminate all canonical momenta. Last, one 
substitutes the integral solutions (\ref{44}), (\ref{46}), and 
(\ref{46a}). Contour deformations are not required; use of the 
partial differential equation (\ref{44c}) is sufficient.   

To compute the time evolution of the electric field we repeat the 
program described above for (\ref{12}). This brings us to 
\begin{equation} 
\hspace{-0mm}
{\bf E}({\bf r},t) = - \int d{\bf r}' \int_{C_3} 
\frac{dz}{2\pi i\epsilon_0} ze^{-izt} 
{\sf G}(z;{\bf r},{\bf r}')\cdot {\bf j} (z;{\bf r}'). \label{47}
\end{equation}
Combination of (\ref{46}) and (\ref{47}) gives for the electric 
displacement 
\[
\hspace{-0mm} 
{\bf D}({\bf r},t)= \epsilon_0 {\bf E}({\bf r},t) - \alpha ({\bf r}) 
{\bf Q}_0({\bf r},t) = {\bf D}({\bf r},0) + \int_{C_3} 
\frac{dz}{2\pi i} \frac{e^{-izt}}{z} {c^2\bf j}(z;{\bf r}) 
\]
\begin{equation} 
\hspace{-0mm}
- \int d{\bf r}' \int_{C_3} 
\frac{dz}{2\pi i} z\varepsilon (z;{\bf r}) e^{-izt}
{\sf G}(z;{\bf r},{\bf r}')\cdot {\bf j} (z;{\bf r}'). \label{48}
\end{equation}
From (\ref{47}) and (\ref{48}) we conclude that one may indeed regard 
$\varepsilon (z;{\bf r})$ as a dielectric function. With the help of 
(\ref{18}) one can clarify the status of 
the local contribution to the electric displacement. It guarantees 
that the field ${\bf D}_L({\bf r},t)$ does not exist, as prescribed by 
Maxwell's equations. Indeed, if one takes the divergence of (\ref{48}) 
and inserts (\ref{44c}), the right-hand side reduces to the divergence 
of ${\bf D}({\bf r},0)$, which equals zero. 

In \cite{HB:1992} the electric field and electric displacement 
were calculated for the case of a spatially homogeneous dielectric, 
coupled to an uncountable number of harmonic oscillators. We can extend 
these results to the case of an inhomogeneous dielectric by considering 
(\ref{47}) and (\ref{48}) for a reservoir, the eigenfrequencies of 
which make up a dense set. This will be done in the next subsection.  

\subsection{Continuum limit} 

For finite $N$ the solution of the extended Ullersma model describes 
reversible dynamics. Therefore, the energy exchange between dielectric 
and reservoir is everlasting. On the other hand, for many experiments on 
optical properties of dielectrics, absorption of photons is omnipresent. 
Hence, there is still a gap between the solutions of the previous 
subsection and experiment. Our dielectric medium does exhibit damping 
phenomena if we manage to create an irreversible energy flow into the 
reservoir. For that purpose, the reservoir should possess an uncountable 
number of degrees of freedom. Such a continuum comes into existence upon 
defining   
\begin{eqnarray} 
\Lambda^{1/2}{\bf Q}_n({\bf r},t)   =  {\bf Q}(n/\Lambda ;{\bf r},t) & 
\hspace{1cm} 
\Lambda^{1/2}{\bf P}_n({\bf r},t)   =  {\bf P}(n/\Lambda ;{\bf r},t) 
\nonumber \\ 
\Lambda^{1/2} \beta_n ({\bf r})   =  \beta (n/\Lambda;{\bf r}) & 
\hspace{1cm} 
\omega_n({\bf r})  =  n/\Lambda \label{49} 
\end{eqnarray} 
for $n=1,2,3,\ldots,N$, and taking the limit 
$\Lambda, N\rightarrow\infty$. Since $n$ becomes arbitrarily large, the 
ratio $n/\Lambda$ may be treated as a continuous variable $\omega$. Note 
that the spatial dependence of $\omega_n$ does not leave any traces. The 
subscript $c$ indicates that the continuum limit is taken. 

We assume that for all nonnegative $\omega$ the function $\beta (\omega)$ 
is regular and smooth. If it decays faster than $1/\sqrt{\omega}$ for 
large $\omega$, then the definition below (\ref{7}) implies that the 
continuum limit of $\tilde{\omega}_0$ exists. The analytic properties of 
the function  
\begin{equation} 
h_c (z) = z^2 -\tilde{\omega}_{0,c}^2 
+\int_0^{\infty} d\omega 
\frac{\omega^2 \beta^2 (\omega)}{\rho^2 (\omega^2 -z^2)} \label{50} 
\end{equation}
radically differ from those of $h(z)$, given in (\ref{14}). The zeros of 
$h(z)$ have united so as to generate a branch cut on the real axis. 
The same mechanism is witnessed for the dielectric function 
$\varepsilon_c (z)$ and the eigenvalue $\lambda_c (k,z)$. In appendix A 
we demonstrate that all of the afore-mentioned functions are analytic 
and nonzero outside the real axis, that is to say, both on the branch 
in the upper half plane and on the branch in the lower half plane. We 
assume that the analytic properties of the eigenvector ${\bf u}(k,z)$ 
are not affected by the continuum limit. 

Now we are well prepared to find out how the solution of the extended 
Ullersma model, which was derived in the previous subsection, behaves 
under the continuum limit. We focus on the electric field, specified in 
(\ref{47}). Treatment of the electric displacement and other fields goes 
by the same methodology. As mentioned earlier, the contour $C_3$ is 
composed of two straight lines enclosing the real axis. We parametrise 
$C_3$ as $\int dz f(z) =\int_{-\infty}^{\infty}d\omega [f(\omega -i\eta) 
-f(\omega + i\eta)]$, where $\eta$ is infinitesimally positive. Next, we 
simplify the integrand with the help of the symmetry 
$[{\sf G}(z^{\ast};{\bf r},{\bf r}')]^{\ast}=
{\sf G}(z;{\bf r},{\bf r}')$. Application of rule (\ref{49}) then leads 
to  
\[
\hspace{-15mm}
{\bf E}({\bf r},t) = \int d{\bf r}' \int_{-\infty}^{\infty} 
\frac{d\omega}{\pi \epsilon_0} e^{-i\omega t} \omega \mbox{Im} 
\left[ {\sf G}_c(\omega +i\eta ;{\bf r},{\bf r}')\right]\cdot 
{\bf j}_{EM}(\omega ;{\bf r}') 
\]
\[ 
\hspace{-2mm}
- \int d{\bf r}' \int_{-\infty}^{\infty} 
\frac{d\omega}{\pi i\epsilon_0} e^{-i\omega t} \omega \alpha ({\bf r}') 
\mbox{Im} \left[
\frac{{\sf G}_c(\omega +i\eta ;{\bf r},{\bf r}')}
{h_c(\omega +i\eta ;{\bf r}')}\right]\cdot {\bf j}_{D}(\omega ;{\bf r}') 
\]
\[
\hspace{-2mm}  
+ \int d{\bf r}' \int_{-\infty}^{\infty} 
\frac{d\omega}{\pi \epsilon_0} \int_0^{\infty} d\omega' e^{-i\omega t} 
\frac{\omega^2 \alpha ({\bf r}')\beta (\omega';{\bf r}')}
{\rho ({\bf r}')} 
\]
\begin{equation} 
\hspace{-2mm} 
\times \mbox{Im} \left\{
\frac{{\sf G}_c(\omega +i\eta ;{\bf r},{\bf r}')}
{h_c(\omega +i\eta ;{\bf r}')[(\omega +i\eta)^2-\omega'^2]}\right\}
\cdot {\bf j}_{R}(\omega ,\omega' ;{\bf r}') . \label{90}
\end{equation} 
The continuum Green function ${\sf G}_c(\omega +i\eta ; {\bf r},{\bf r}')$ 
is determined by (\ref{44b}), with the replacements $z\rightarrow \omega 
+i\eta$ and $\lambda \rightarrow \lambda_c$ carried out. The new  
source vectors are given by 
\begin{eqnarray} 
c^2{\bf j}_{EM} (\omega ;{\bf r}) & = & 
-i\epsilon_0 \omega {\bf A}({\bf r},0) + {\bf \Pi}({\bf r},0) 
+ \left[ \alpha ({\bf r}) {\bf Q}_0({\bf r},0)\right]_T \nonumber \\ 
c^2{\bf j}_D (\omega ;{\bf r}) & = & 
i\omega^2 {\bf Q}_0 ({\bf r},0) -\frac{\omega}{\rho ({\bf r})} 
{\bf P}_0({\bf r},0) \nonumber \\ 
c^2{\bf j}_R(\omega ,\omega' ;{\bf r}) & = & i\omega'^2 
{\bf Q}(\omega' ;{\bf r},0)- \frac{\omega}{\rho ({\bf r})} 
{\bf P}(\omega' ;{\bf r},0). \label{91} 
\end{eqnarray} 
In \cite{SW:2004} the above solution is derived on the basis of Laplace 
transformation. In that paper, the reservoir contains a 
continuum of oscillators right from the beginning. 

By taking the continuum limit as demonstrated above, we fully extend the 
results of \cite{HB:1992} to the case of an inhomogeneous 
dielectric. However, there exists an alternative manner to implement the 
continuum limit. Instead of keeping the upper and lower part of $C_3$ 
together, one can decide to sever these parts from each other. The 
ensuing representations are useful to analyse how fields behave for long 
times. We keep the time strictly positive and focus again on the electric 
field. For some integrands figuring in (\ref{47}) it is wise to perform 
first the substitution $ z/[f(z)] = [z^2 -f(z)]/[z f(z)]+1/z$, where 
$f(z)$ stands for $\lambda_c(k,z)$ or $h_c(z)$. One then isolates the 
term $1/z$, which decays slowly as $|z|$ becomes large. The convergence 
of the remaining integrand improves.  

The lower part of $C_3$ can be closed by means of the arc 
$z=R\exp (i\phi)$, with $-\pi \le \phi \le 0$. Owing to the choice $t>0$, 
the exponential factor $\exp (-izt)$ causes the integrand to shrink to 
zero on the arc. In the interior of the closed contour the integrand is 
analytic, so the lower part of $C_3$ does not make any contribution. The 
upper part of $C_3$ yields an integral over all real frequencies $\omega$, 
which is convergent in virtue of the above substitution. Of course, for 
the term $1/z$ one cannot treat the upper and lower parts of $C_3$ 
separately; instead, the residue theorem offers a way out. 

If we execute the above instructions for the electric field, then the 
following expression emerges: 
\[ 
\hspace{-3mm}
{\bf E}({\bf r},t) = {\bf E}({\bf r},0)+\int_{-\infty}^{\infty} 
\frac{d\omega}{2\pi i\epsilon_0}e^{-i\omega t} 
\frac{\alpha ({\bf r})[h_c(\omega +i\eta; {\bf r}) -\omega^2]} 
{(\omega +i\eta) h_c(\omega + i\eta;{\bf r})} 
{\bf Q}_0({\bf r},0) 
\]
\[
\hspace{10mm} 
+ \int d{\bf r}'\int_{-\infty}^{\infty} 
\frac{d\omega}{2\pi i\epsilon_0} e^{-i\omega t} 
\left[\omega{\sf G}_c(\omega +i\eta ;{\bf r},{\bf r}')
-\frac{c^2}{\omega +i\eta}
\mbox{{\boldmath $\delta$}} ({\bf r}-{\bf r}')\right]  
\]
\[
\hspace{10mm} \cdot  
\left[{\bf j}_{EM}(\omega ;{\bf r}') - 
\frac{\omega^2 \alpha ({\bf r}'){\bf Q}_0({\bf r}',0)}
{c^2h_c(\omega + i\eta ;{\bf r}')} \right]   
\]
\[
\hspace{10mm} 
+ \int d{\bf r}'\int_{-\infty}^{\infty} 
\frac{d\omega}{2\pi i\epsilon_0} e^{-i\omega t} 
\frac{\omega^2 \alpha({\bf r}')} 
{\rho ({\bf r}') h_c(\omega +i\eta ;{\bf r}')}  
{\sf G}_c(\omega +i\eta ;{\bf r},{\bf r}') 
\]
\begin{equation} 
\hspace{10mm} 
\cdot \left[ -\frac{i}{c^2}
{\bf P}_0({\bf r}',0) +\int_0^{\infty}d\omega' 
\frac{\beta (\omega';{\bf r}')}
{(\omega +i\eta )^2 -\omega'^2} {\bf j}_R(\omega,\omega';{\bf r}') 
\right]  .
\label{52} 
\end{equation} 
As before, $\eta$ is infinitesimally positive. In computing 
$h_c(\omega +i\eta;{\bf r})$ the upper branch of the function 
(\ref{50}) must be used. Consequently, each integrand of (\ref{52}) is 
analytic in the upper half of the complex $\omega$ plane. For $t=0$ one 
may then replace the contour by an arc $\omega = R\exp (i\phi)$, with 
$0\le \phi\le\pi$. On account of (\ref{14}), (\ref{23}) and (\ref{16}), 
each integrand decays as $1/R^2$ or faster, so there are no integrals 
surviving the choice $t=0$. In the continuum limit the initial 
condition for the electric field is still satisfied. 

To uncover irreversible behaviour, we have to analyse (\ref{52}) for 
large times. If the symmetry relation 
\begin{equation} 
\left[\beta^2(-z^{\ast})\right]^{\ast} = \beta^2 (z)  \label{53} 
\end{equation} 
holds true, and $\beta (z)$ is well-behaved at the origin, then the 
analytic continuation of $\varepsilon_c (\omega +i\eta)$ and 
$\lambda_c(k,\omega +i\eta)$ can be effectuated. Thus the infinitesimal 
increment $i\eta$ of the argument $\omega$ becomes redundant. From 
(\ref{19}) and the continuation of $h_c(\omega +i\eta)$ we obtain 
the symmetry relation $[\varepsilon_c (-\omega^{\ast})]^{\ast} = 
\varepsilon_c (\omega )$, which is in accordance with classical theory 
of the dielectric constant \cite{JAC:1975}. All this is shown in 
appendix A. There we also prove that $\varepsilon_c (\omega)$ is 
analytic and different from zero for all $\mbox{Im}\omega \ge 0$. 
Incidentally, if the symmetry (\ref{53}) is absent, indeed circumstances 
exist under which the process of analytic continuation fails 
\cite{HAA:1985}. 
  
Now everything is ready to shift in (\ref{52}) the contour into the 
lower half plane. The poles at $\omega = -i\eta$ yield residues that 
erase the initial condition for the electric field.  We adapt $\gamma$ 
such that all integrands accompanying ${\bf A}$, ${\bf \Pi}$, ${\bf Q}_0$, 
or ${\bf P}_0$, are analytic in the strip $-\gamma \le \mbox{Im}\omega 
\le 0$. Then one may integrate along the line $\mbox{Im}\omega = -\gamma$
instead of the real axis. The parameter $\gamma$ surely differs from zero, 
otherwise the above process of analytic continuation would fail. The 
exponential $\exp (-i\omega t)$ produces a factor of $\exp (-\gamma t)$, 
which induces absorption in the dielectric, as desired. By comparing 
(\ref{47}) and (\ref{52}) one can trace the origin of the irreversible 
behaviour. It resides in the fact that, as a result of the continuum 
limit, the analytic properties of $h(z)$ undergo a radical change.  

Use of (\ref{19}) in (\ref{52}) gives a dielectric function 
$\varepsilon_c (\omega )$ that is analytic in the upper half plane. 
Therefore, irreversibility for $t\rightarrow \infty$ is coupled to 
causality for the dielectric function. Indeed, if one takes the continuum 
limit for $t$ negative, one has to operate on the lower branch of 
(\ref{50}). Then the dielectric function is analytic in the lower half of 
the complex $\omega$ plane. One encounters the combination of 
irreversibility for $t\rightarrow -\infty$ and  anti-causality for the 
dielectric function. Of course, these conclusions do not depend on 
(\ref{53}). In absence of this symmetry one may resort to arguments of 
Riemann-Lebesgue type. 

The terms of (\ref{52}) that contain the canonical fields of the reservoir 
have not yet been considered. The corresponding integrand has poles at 
$\omega = \pm\omega' -i\eta$, the residues of which generate oscillating 
contributions. Upon using the symmetry relations for ${\sf G}_c$ and $h_c$, 
we arrive at the following asymptotic expression for the electric field, 
valid for large times: 
\[ 
\hspace{-25mm} 
{\bf E}({\bf r},t) \sim 
\]
\begin{equation} 
\hspace{-25mm} 
-\int d{\bf r}' \int_0^{\infty}d \omega
 e^{-i\omega t -i\phi} \left\{\frac{\omega \rho ({\bf r}')\mbox{Im}
\left[\varepsilon_c (\omega ;{\bf r}')\right]}{2\pi\epsilon_0}\right\}^{1/2} 
{\sf G}_c(\omega ;{\bf r},{\bf r}')\cdot {\bf j}_R(\omega,\omega ;{\bf r}') 
+\mbox{hc}.  
\label{95}
\end{equation}   
The phase $\phi = \arg(h_c)$ results from eliminating the coupling 
parameters $\alpha$ and $\beta$ in favour of the dielectric function. To do 
this, the usual distributional calculus should be applied to (\ref{50}). The 
imaginary part of the integrand on the right-hand side of (\ref{50}) is 
proportional to a Dirac delta function. 

In virtue of a Riemann-Lebesgue argument, the expectation value of the 
integral of (\ref{95}) vanishes for large times. Hence, the same is true for 
the electric-field operator ${\bf E}({\bf r},t)$. Still, the oscillatory 
contributions of the reservoir qualitatively differ from those exhibiting 
exponential damping. The oscillations bring about quantum fluctuations that 
do not die out for large times. To make this explicit, we model the initial 
correlations in the reservoir as 
\begin{equation} 
\left\langle {\bf P}(\omega ;{\bf r},0){\bf P}(\omega';{\bf r}',0)
\right\rangle = \frac{1}{3} 
\left\langle {\bf P}^2 (\omega;{\bf r},0)\right\rangle 
\delta (\omega -\omega') \mbox{{\boldmath $\delta$}} ({\bf r}-{\bf r}') 
\label{56}   
\end{equation} 
where the last {\boldmath $\delta$} function is a tensor. The brackets 
indicate that an expectation value is taken. For the auto-correlations 
of ${\bf Q}(\omega ;{\bf r},0)$ the above model is assumed as well. 
Cross-correlations of ${\bf P}(\omega ;{\bf r},0)$ and 
${\bf Q}(\omega ;{\bf r},0)$ are discarded. From the electric field we 
construct a quadratic form, take the expectation value, and process 
the result with the help of (\ref{95}) and (\ref{56}). Utilising again 
a Riemann-Lebesgue argument and introducing the energy density of the 
reservoir as 
\begin{equation} 
{\cal H}_R (\omega ;{\bf r}) = 
\frac{1}{2} \rho^{-1}({\bf r}) {\bf P}^2(\omega ;{\bf r},0) + 
\frac{1}{2} \omega^2 \rho ({\bf r}) {\bf Q}^2 (\omega ;{\bf r},0) 
\label{58}
\end{equation} 
we find the asymptotic result  
\[
\hspace{-20mm} 
\lim_{t\rightarrow\infty} \left\langle {\bf E}({\bf r},t)  
{\bf E}({\bf r}',t)\right\rangle =  
\]
\begin{equation} 
\hspace{-20mm} 
 \int d{\bf r}''  
\int_0^{\infty} \frac{d \omega}{3\pi \epsilon_0 c^4} \omega^3  
{\sf G}_c(\omega ;{\bf r},{\bf r}'')\cdot 
{\sf G}_c^{\ast}(\omega ;{\bf r}'',{\bf r}') 
\mbox{Im} \left[\varepsilon_c (\omega ;{\bf r}'')\right] 
\left\langle {\cal H}_R (\omega ;{\bf r}'')\right\rangle +\mbox{cc}. 
\label{57} 
\end{equation} 
In the continuum limit any expectation value of the electric field 
will eventually decay to zero, but the quantum fluctuations stay 
alive. They are fuelled by the energy that is available in the 
reservoir.    

\section{Summary and conclusion} 

The quantisation of the electromagnetic field in the presence of 
nonrelativistic macroscopic matter constitutes a problem of 
formidable magnitude. In principle, any serious treatment should 
start from the minimal-coupling Hamiltonian that takes into account 
all electromagnetic interactions at the atomic level. It needs no 
argument that such a rigorous approach is scarcely possible, unless 
one is prepared to deploy heavy numerical means as soon as it comes to 
predicting experimental findings. However, one then misses the 
opportunity to see what kind of mathematical mechanisms are at work 
behind such processes as spontaneous emission or scattering of photons. 
Therefore, already during the early years of the subject people were 
looking for shortcuts so as to obtain concise and transparent theories 
\cite{JAU:1948}. Nowadays a generally accepted quantisation scheme of 
phenomenological nature is available. 

Because of their experimental relevance, rules for QED in absorptive 
matter deserve to be put on a firm foundation. This idea was pursued 
by Huttner and Barnett \cite{HB:1992}. Sacrificing the direct contact 
with the atomic level, they solved a so-called damped-polariton model. 
Their microscopic expression for the quantised electromagnetic field 
in an absorptive dielectric initiated a lot of activity on the 
construction of phenomenological quantisation rules. Despite this 
progress, a for experimentalists most important issue is still 
pending, namely, the microscopic quantisation in a dielectric with 
both spatial inhomogeneities and losses. In addition to that, it is  
unclear how the damped-polariton model relates to the old oscillator 
models for dissipative quantum dynamics that were proposed in the 
1960s \cite{SEN:1960}--\cite{DEK:1981}.      

Motivated by the last remarks, we exchange in this work the continuum 
of the damped-polariton model for a reservoir that consists of a 
finite number of independent harmonic oscillators. We thus consider 
the Ullersma configuration \cite{ULL:1966}, extended with an 
electromagnetic sector. For all parameters a spatial dependence is 
admitted, so that the oscillator model at hand is capable of 
describing spatial inhomogeneities. To give the reader the chance to 
make a direct comparison between the quantum and the classical 
evolution of the model, we first solve Hamilton's equations for the 
classical canonical fields. The latter are expressed as sums of 
independent normal modes. By associating with each normal mode a 
harmonic oscillator, we can invoke Dirac's method for the 
quantisation of the dynamics. The corresponding ladder operators can 
be employed to cast the Hamiltonian into diagonal form. Hence, the 
time evolution of all fields can be made explicit. Having derived the 
solution of the extended Ullersma model, we can identify the 
inhomogeneous dielectric function. Also, we can meet one of our prime 
objectives, the generalisation of the results of Huttner and Barnett 
\cite{HB:1992} to the case of an inhomogeneous dielectric. This job 
is completed by subjecting our discrete reservoir to a continuum 
limit, which makes the coupling parameter of each separate oscillator 
infinitely weak, and at the same time converts the collection of 
eigenfrequencies into a dense set.    

The discrete character of the Ullersma reservoir brings us important  
advantages. We can completely avoid the mathematical complexities that 
accompany the use of distributions. For example, we need not adapt 
the solution of the evolution equations through addition of an unknown 
term, containing a delta function \cite{HB:1992,FAN:1961,SWO:2004}. To 
derive the results of this paper only simple technical tools are 
required. We apply linear algebra to prove orthogonality and 
completeness for the eigenvectors of operator $L$, which governs the 
spatial structure of the normal modes. Hence, these eigenvectors are 
the natural candidates for composing the Green function, as becomes 
manifest when computing the solutions for the canonical fields. The 
other standard ingredient we call in, is the residue theorem. It 
allows us to reduce all discrete sums of normal modes to complex 
integrals. The ensuing contour $C_1$ is rather awkward, as it is 
composed of separate circles around poles on the real axis. 
Fortunately, $C_1$ gives way to frequency integrals upon performing 
contour deformation. In doing so, we generate correction terms of local 
nature, in which the Green function no longer figures. These terms can 
be interpreted as Langevin noise operators \cite{HB:1992}. The local 
correction to the electric displacement ensures that the divergence of 
the latter equals zero, in accordance with Maxwell's equations. The easy 
access to the mathematical structure of our solutions enables us to 
scrutinise their behaviour under the continuum limit. The function $h$ 
plays a central role. Its poles gradually cover the real axis, and 
finally give rise to a branch cut. The dielectric function is analytic 
on both sides of the cut, but the sign of the time $t$ determines on 
which branch the fields differ from zero. This leads to the conclusion 
that (anti)causality of the dielectric function is intimately linked to 
irreversibility of the dynamics for $t$ approaching (minus) infinity.     

A further advantage offered by the Ullersma reservoir is the connection 
with recent work on decoherence and entanglement. Several papers  
describe how a reservoir of the Ullersma type can bring about 
decoherence of quantum superpositions and entanglement between qubits 
\cite{ROM:1997}--\cite{BRA:2002}. Results on these processes 
are valuable with an eye to quantum computing and other future 
applications. Inevitably, in any real device electromagnetic fields will 
participate in the interactions. Therefore, it would be worthwhile to 
study the reservoir-induced quantum phenomena in the presence of an 
electromagnetic sector. One would like to assess the influence of the 
quantum noise that is produced by the electromagnetic field. As we saw 
in this work, the energy density of the reservoir fuels the quantum 
fluctuations of the electric field. They outlive any decoherence or 
entanglement. The possible fragility of these quantum processes can 
surely be investigated with the aid of the extended Ullersma model. 
Especially the influence of temperature must be critically regarded.  

\appendix
\section{Results for $h$ and related functions}

In the main text we exploited some analytic properties of the functions 
$h(z;{\bf r})$, $\varepsilon (z;{\bf r})$, and $\lambda (k,z)$. For the  
first two functions, all properties can be proved by means of simple 
techniques, as we shall discuss below. When we turn to $\lambda (k,z)$, 
however, the going becomes heavier, due to the fact that the partial 
differential equation (\ref{16}) intervenes. We shall need a few 
plausibility arguments in order to make progress. A rigorous 
mathematical investigation of (\ref{16}) lies outside the scope of this 
paper.   

The function $h(z;{\bf r})$, defined in (\ref{14}), is equal to the 
ratio of two finite polynomials in $z$. Hence, $h^{-1}(z;{\bf r})$ is a 
meromorphic function of $z$, the poles of which follow by solving the 
equation $h(z;{\bf r})=0$. The imaginary part of $h(z;{\bf r})$ is 
given by 
\begin{equation} 
\mbox{Im} h(z;{\bf r}) = 
\mbox{Im}(z^2)\left[ 1+ \sum_{n=1}^N 
\frac{\beta_n^2 ({\bf r}) \omega_n^2({\bf r})}{\rho^2 ({\bf r}) 
|z^2 -\omega_n^2({\bf r})|^2}\right]  . 
\label{A1} 
\end{equation} 
From this result the inequality 
\begin{equation} 
\left| \mbox{Im} h(z;{\bf r})\right| > \left| \mbox{Im}( z^2) \right| 
\label{A2} 
\end{equation} 
is manifest. We see that the square $z^2$ must be real, otherwise 
$h(z;{\bf r})$ cannot equal zero. For $z^2 \le 0$ the inequality  
\begin{equation} 
h(ib;{\bf r}) < -\omega_0^2({\bf r}) \label{A3} 
\end{equation} 
comes into play, where $b$ is real. Clearly, $h(z;{\bf r})$ can equal 
zero only if $z^2$ is real and positive. We confirm our surmise that  
$h^{-1}(z;{\bf r})$ is meromorphic, with all poles lying on the real 
axis, symmetric with respect to the origin. 

The function $h_c(z;{\bf r})$, the continuum counterpart of 
$h(z;{\bf r})$, is specified in (\ref{50}). As long as $\mbox{Im}z$ 
differs from zero, the factor of $(\omega^2-z^2)$ cannot render the 
denominator of (\ref{50}) zero. This guarantees that $h_c(z;{\bf r})$ is 
analytic outside the real axis, i.e., on each of its two branches. Since 
$h_c(z;{\bf r})$ is obtained by applying a limiting process to 
$h(z;{\bf r})$, the inequalities (\ref{A2}) and (\ref{A3}) are valid for 
$h_c(z;{\bf r})$ as well. Therefore, $h_c(z;{\bf r})$ does not vanish 
outside the real axis.  

In virtue of (\ref{19}) the dielectric function 
$\varepsilon_c(z;{\bf r})$ inherits the branch cut of $h_c(z;{\bf r})$, 
and is analytic on each of its two branches. The equality 
$\varepsilon_c (z;{\bf r})=0$ implies that $\mbox{Im} h_c(z;{\bf r})$ 
becomes zero. For $\mbox{Re}z\ne 0$ this contradicts (\ref{A2}). As  
before, the case $\mbox{Re}z=0$ must be checked separately. This can 
happen via (\ref{A3}). Altogether, we conclude that 
$\varepsilon_c(z;{\bf r})$ is analytic and nonzero on each of its two 
branches.      

Before we can make statements about the eigenvalue $\lambda (k,z)$, 
we have to pay attention to the eigenvector ${\bf u}(k,z;{\bf r})$, 
which is the solution of (\ref{40}) and (\ref{16}). As is manifest from 
(\ref{18}), the dependence of the operator $L(s;{\bf r})$ on the real 
variable $s$ is smooth, except for the points where $h(s;{\bf r})$ equals 
zero. The corresponding divergencies do not appear in 
${\bf u}(k,s;{\bf r})$, owing to the normalisation (\ref{21}). This 
incites us to suppose that ${\bf u}(k,s;{\bf r})$ is regular and smooth 
for all real $s$. Consequently, ${\bf u}(k,z;{\bf r})$ is analytic in a 
certain strip around the real axis. Outside this strip the operator 
$L(z;{\bf r})$ is nowhere singular, so it seems reasonable to assume 
that ${\bf u}(k,z;{\bf r})$ is an entire function in the complex $z$ 
plane. Then the same goes for $[{\bf u}(k,z^{\ast};{\bf r})]^{\ast}$, 
by the rules of complex conjugation.  

The results for $h_c(z;{\bf r})$ and $\varepsilon_c(z;{\bf r})$, which 
were proved above, illustrate that the continuum limit does not create 
any new singularities outside the real axis. It merely modifies the 
character of the singularities that already exist on the real axis. 
Poles unite and form a branch cut. We assume that also for 
${\bf u}(k,z;{\bf r})$ the continuum limit does not create any new 
singularities. Said differently, the eigenvector remains an entire 
function. Note that all of our assertions on ${\bf u}(k,z;{\bf r})$ 
are trivial for the case of a homogeneous dielectric, because then 
the dependence of ${\bf u}(k,z;{\bf r})$ on $z$ is absent.  

In (\ref{24}) the eigenvalue $\lambda (k,z)$ is expressed in terms of a 
spatial integral over the volume $V$. The above results tell us that the 
integrand is analytic in the whole complex $z$ plane, except for the 
real axis. Since $V$ is finite, the integral is convergent if the 
integrand is bounded. We therefore may conclude that $\lambda (k,z)$ is 
analytic as long as $\mbox{Im}z$ differs from zero. The foregoing 
argument can be repeated for $\lambda_c (k,z)$, so this function is 
analytic on both of its two branches. 

The identity $[\lambda (k,z^{\ast})]^{\ast}= \lambda (k,z)$ allows us 
to put forward 
\begin{equation} 
\mbox{Im} \lambda (k,z) = \mbox{Im} z \left[ 
\frac{\lambda (k,z) - \lambda (k,z^{\ast})}{z-z^{\ast}} \right]. 
\label{A4} 
\end{equation} 
Let us make $\mbox{Im}z$ small, and choose $\mbox{Re}z$ such that 
$h(z;{\bf r})$ is nonzero for $\mbox{Im}z=0$. We then avoid 
divergencies on the real axis. To elaborate (\ref{A4}) we invoke the 
identity 
\begin{equation} 
\hspace{-25mm} 
\frac{d\lambda (k,z)}{dz} 
\int d{\bf r} \left[ 
{\bf u}(k,z^{\ast};{\bf r})\right]^{\ast}\cdot {\bf u}(k,z;{\bf r}) 
 = \int d{\bf r} \left[ 
{\bf u}(k,z^{\ast};{\bf r})\right]^{\ast}\cdot {\bf u}(k,z;{\bf r}) 
\frac{dz^2\varepsilon (z;{\bf r})}{dz}. \label{A6}
\end{equation} 
It can be proved with the help of (\ref{24}) and partial integration. 
Before carrying out the replacement $s\rightarrow z$ in (\ref{24}), one 
has to multiply the left-hand side by the norm $\langle {\bf u}(k,s), 
{\bf u}(k,s)\rangle$. After use of (\ref{A6}) the identity (\ref{A4}) 
attains the form 
\begin{equation} 
\mbox{Im} \lambda (k,z) \approx \mbox{Im}(z^2) \left[ \int d{\bf r} 
\left| {\bf u}(k,z;{\bf r})\right|^2 
\frac{dz^2 \varepsilon (z;{\bf r})}{dz^2} 
\right]_{{\mbox{\scriptsize Im}}z=0}   
\label{A5} 
\end{equation}        
where $\mbox{Im}z$ is small and $h(z;{\bf r})$ must be nonzero for 
$\mbox{Im}z=0$.  

The result (\ref{A5}) invites us to employ the inequality 
\begin{equation} 
\frac{ds^2 \varepsilon (s;{\bf r})}{ds^2} > 1 \label{A7} 
\end{equation} 
which follows from (\ref{19}). This brings us to 
\begin{equation} 
\left| \mbox{Im} \lambda(k,z)\right| > 
\left| \mbox{Im}( z^2) \right| 
\label{A8} 
\end{equation} 
where $\mbox{Im}z$ is small and $\mbox{Re}z$ is nonzero. Hence, the 
function $\lambda (k,z)$ surely differs from zero in the vicinity of 
the real axis. Note that the case $\mbox{Re}z=0$ is unimportant, 
because (\ref{40}) and (\ref{24}) force $\lambda (k,z)$ to be real 
and negative on the imaginary axis. By taking the continuum limit of 
$\lambda (k,z)$, we do not create zeros for small $\mbox{Im}z$, 
because the lower bound (\ref{A8}) does not depend on $N$ and 
$\Lambda$. Hence, close to the real axis $\lambda_c(k,z)$ is nonzero.   

The above findings on analyticity and location of zeros permit us to 
make a preliminary statement. We may claim that outside the real axis 
$\lambda^{-1}(k,z)$ is analytic whenever $\lambda (k,z)$ differs from 
zero. Therefore, in the vicinity of the real axis $\lambda^{-1}(k,z)$ 
is analytic. Now the question arises of what happens further out in 
the complex plane, and on the real axis itself. 

For the case of a homogeneous dielectric $\lambda (k,z)$ has $(2N+4)$ 
zeros and $(2N+2)$ poles. All of these are located on the real axis, 
as follows from (\ref{A4}). Under a smooth transition from a 
homogeneous to an inhomogeneous dielectric, the $(4N+6)$ zeros and 
poles cannot leave the real axis. This has been demonstrated above. 
What we now assume is that the transition does not generate any new 
zeros or any new poles. Then $\lambda^{-1}(k,z)$ is meromorphic, with 
all poles located on the real axis. Moreover, $\lambda_c(k,z)$ will 
not possess any zeros on each of its two branches. 
    
To carry out a check on our assumption, we utilise the 
argument principle \cite{CON:1978}. For $\lambda (k,z)$ it can be 
formulated as  
\begin{equation} 
\int_{C_2} \frac{dz}{2\pi i} 
\frac{ d\lambda (k,z)/dz}{\lambda (k,z)} = (2N+4)-(2N+2). \label{A9} 
\end{equation}
On the right-hand side the number of poles of $\lambda (k,z)$ is 
subtracted from the number of zeros of $\lambda (k,z)$.  
The contour $C_2$ is a large circle centered around the origin. To 
prove (\ref{A9}) we appeal to (\ref{A6}) once more. The ensuing 
integrand can be simplified with the help of (\ref{B1}) and the fact 
that $\varepsilon (z;{\bf r})$ converges to unity for $|z|$ large. 
Then (\ref{A9}) boils down to the condition $\int_{C_2} dz /
(2\pi iz)= 1$. This elementary integral completes our consistency 
check. 

Our last job is the analytic continuation of $h_c(z)$. The symmetry 
(\ref{53}) allows us to extend the contour of (\ref{50}) to the 
negative real axis. By assumption, $\beta^2 (z)$ is regular and 
smooth for all real $z$, so it must be analytic inside a certain 
strip around the real axis. Keeping $\mbox{Im}z$ positive as required 
by (\ref{52}), we shift in (\ref{50}) the contour downwards, until it 
runs below the pole at $\omega = -z$. Upon evaluating the residue we 
acquire an alternative form of (\ref{50}), given by 
\begin{equation} 
h_c(z) = z^2 -\tilde{\omega}_{0,c}^2 + \frac{i\pi z}{2\rho^2} 
\beta^2 (-z) + \frac{1}{2}\int_{C_4} d\omega 
\frac{\omega^2\beta^2(\omega)}{\rho^2 (\omega^2-z^2)}. \label{A20} 
\end{equation} 
The contour $C_4$ is a straight line, running parallel to the real 
axis at $\mbox{Im}\omega = -\gamma$, with $\gamma $ positive. 

Rather than the upper half plane, the strip $-\gamma < \mbox{Im}z 
< \gamma$ constitutes the region where the representation (\ref{A20}) 
is analytic. We thus have succeeded in finding an analytic continuation 
of (\ref{50}) below the real axis. From (\ref{A20}) we infer the 
symmetry relation 
\begin{equation} 
\left[h_c (-z^{\ast})\right]^{\ast} = h_c (z). \label{A21} 
\end{equation} 
Note that the operation $z\rightarrow -z^{\ast}$ maps the strip 
$-\gamma < \mbox{Im}z < \gamma $ onto itself, so in employing 
(\ref{A21}) domain questions do not arise. In view of (\ref{19}) and 
(\ref{24}), as  well as the fact that ${\bf u}(k,z)$ is an entire 
function, the analytic continuation of $\varepsilon_c (\omega +i\eta)$ 
and $\lambda_c (k,\omega +i\eta)$ can be performed on the basis of 
(\ref{A20}). 

All properties of $h$ and related functions, which are needed in the 
main text, have now been proved or made plausible. In essence only 
one basic assumption on the partial differential equation (\ref{16}) 
underlies our discussion. For the eigenvector ${\bf u}(k,z;{\bf r})$,  
the eigenvalue $\lambda (k,z)$, as well as its inverse 
$\lambda^{-1} (k,z)$, both the continuum limit, and the transition 
from a homogeneous to an inhomogeneous dielectric, do not give birth 
to singular points out in the complex plane. As a justification, 
we point out that all singular points of the operator $L(z;{\bf r})$ 
lie on the real axis, regardless of the decision to carry out one of 
the two afore-mentioned procedures. Of course, the check (\ref{A9}) 
provides further support for our views. 

\section{Verification of two canonical commutation relations} 

In subsection 3.2 we commenced the verification of the upper two 
commutators (\ref{37}). Upon making the transition to complex 
integration, we arrived at the sufficient conditions 
(\ref{39})--(\ref{42}). The reflection principle 
$[\lambda (k,z^{\ast})]^{\ast} = \lambda (k,z)$ ensures that the 
integrals (\ref{39}) and (\ref{42}) are invariant under complex 
conjugation, combined with the interchange of position vectors and 
tensorial indices. This invariance is imposed by the tensorial delta 
functions figuring in the right-hand sides of (\ref{39}) and (\ref{42}). 
We first get over with (\ref{39}), which is the simplest condition.  

\begin{figure}[h!]
\psfrag{a}{(a)}
\psfrag{b}{(b)}
\psfrag{c}{(c)}
\psfrag{c1}{$C_1$}
\psfrag{c2}{$C_2$}
\psfrag{c3}{$C_3$}
\begin{center}
\includegraphics[height=3cm, width=11cm]{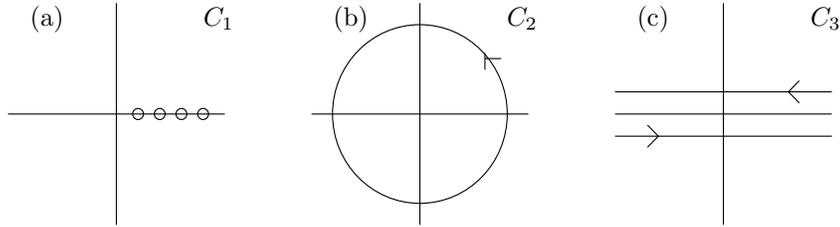}
\end{center}
\caption{Integration contours in the complex plane $z$. Horizontal axis: $\mbox{Im}
z=0$; vertical axis: $\mbox{Re} z=0$. (a) Contour $C_1$ consists of
$(2N+4)$ circles of infinitesimal radius; each circle runs in
counterclockwise direction and is centered around a point of the set $\{z=
\Omega(k,l)\}$. (b) Contour $C_2$ consists of a single circle of large
radius, centered around the origin. (c) Contour $C_3$ consists of two
straight lines running above and below the real axis at infinitesimal distance.}
\end{figure}

With the help of (\ref{40}) and the transformation $z\rightarrow -z$, we
can include the poles at $z= -\Omega (k,l)$ in the contour. In front of the
integral a factor of $1/2$ appears. The new contour $C_2$ (see figure) is
a circle $z=R\exp (i\phi ), 0\le \phi \le 2\pi$. All poles $z=\pm\Omega
(k,l)$ lie in the interior of $C_2$, because $R$ can be taken arbitrarily
large. Note that, except for the poles $z=\pm \Omega (k,l)$, the integrand
is everywhere analytic. A crucial advantage of $C_2$ over $C_1$ is that one
can benefit from asymptotic results. If $|z|$ is large, (\ref{16}) has the
algebraic form
\begin{equation}
\frac{z^2}{\lambda (k,z)} {\bf u}(k,z;{\bf r}) \sim {\bf u}(k,z;{\bf r}). 
\label{B1}
\end{equation} 
The condition (\ref{39}) becomes 
\begin{equation} 
\sum_k \int_{C_2} \frac{dz}{2\pi i} \frac{1}{z} 
[{\bf u}_T(k,z^{\ast};{\bf r}')]^{\ast} 
 {\bf u}_T(k,z;{\bf r})  = 
\mbox{{\boldmath $\delta$}}_T({\bf r}'-{\bf r}). 
\label{B2}
\end{equation}
Use of (\ref{B1}) has modified the analytical structure of the integrand. 
The poles at $z= \pm \Omega (k,l)$ have given way to a single pole at the 
origin. 

We effectuate a last deformation of our integration contour. Let $C_3$ 
enclose the real axis by means of two straight lines running from $+\infty 
+i\eta $ to $-\infty +i\eta$, and from $-\infty -i\eta $ to 
$+\infty -i\eta$, where $\eta$ is infinitesimally positive. The 
integrand in (\ref{B2}) is still analytic outside the real axis, so 
integration along $C_3$ instead of $C_2$ is permitted. Now we are in a 
position to appeal to the completeness relation (\ref{23}). This leaves us 
with the condition $\int dz /(2\pi iz) =1$, which is indeed true for the 
contour $C_3$.         

To get to grips with the analytical structure of the integrand of 
(\ref{42c}), we make the replacement $s\rightarrow z$ in (\ref{16}), and 
write the result as 
\[
\frac{\alpha^2 ({\bf r}) z^2 {\bf u}(k,z;{\bf r})}
{\epsilon_0 \rho ({\bf r}) h(z;{\bf r}) \lambda (k,z)} = 
\]
\begin{equation}
\lambda^{-1}(k,z)  
\left\{ -c^2\nabla \times \left[\nabla \times 
{\bf u}(k,z;{\bf r})\right] + z^2 {\bf u}(k,z;{\bf r})\right\} 
-{\bf u}(k,z;{\bf r}) . \label{B3} 
\end{equation} 
Obviously, the function on the left-hand side has the same analytic 
structure as $\lambda^{-1}(k,z)$. Hence, (\ref{42c}) does not really cause 
any new complications. We first include the regular point $z=0$ in the 
contour $C_1$, and next perform the substitution (\ref{B3}). Deformation 
of contours eventually leads to the condition $\int dz/[z h(z;{\bf r})]=0$, 
where $C_3$ is the contour. As shown in appendix A, all poles of the 
meromorphic function $h^{-1}(z;{\bf r})$ are located on the real axis. 
After deformation of $C_3$ to the large circle $C_2$, the integral indeed 
disappears, because $h(z;{\bf r})$ behaves as $z^2$ for $|z|$ large. 

Since the integrand of (\ref{42}) has two factors of $h$ in the 
denominator, its analytic structure must be carefully investigated before 
any contour deformations can be undertaken. This can be done by adding a 
new contribution to the integrand, the analytic structure of which is 
determined by the function $h^{-1}(z;{\bf r}')$. Instead of the integrand 
itself, we consider the sum    
\begin{equation} 
\hspace{-12mm} 
{\sf f}(k,z;{\bf r}',{\bf r})= 
\frac{z^4 [{\bf u}(k,z^{\ast};{\bf r}')]^{\ast} {\bf u}(k,z;{\bf r})}
{ h(z;{\bf r}')h(z;{\bf r}) \lambda (k,z)}  + 
\frac{z^2\epsilon_0 \rho({\bf r}) 
[{\bf u}(k,z^{\ast};{\bf r}')]^{\ast} {\bf u}(k,z;{\bf r})}
{ \alpha^2 ({\bf r}) h(z;{\bf r}') }. \label{B4}
\end{equation}
The reason is that the poles of the function ${\sf f}$ can be easily located. 
Upon employing (\ref{B3}) twice, one recognises that ${\sf f}$ has the same 
analytic structure as $\lambda^{-1} (k,z)$. Thus the integrand of (\ref{42}) 
has poles both for $\lambda (k,z)=0$ and for $h(z;{\bf r}')=0$. Again, we 
include the regular point $z=0$ in the contour $C_1$.  
Now we can start shifting contours. At each point $z= \Omega (k,l)$ the 
second contribution to ${\sf f}$ is regular, so integration along $C_1$ 
yields zero. Consequently, in (\ref{42}) the integrand may be exchanged for 
${\sf f}/z$. Next, one can replace $C_1$ by $C_2$, if the transformation 
$z\rightarrow -z$ is used. Upon employing (\ref{B4}) once more, one 
arrives at the condition 
\begin{equation} 
\hspace{-25mm}
\sum_k \int_{C_2} \frac{dz}{2\pi i}  \left[ 
\frac{z^2}{h(z;{\bf r})\lambda (k,z)} + 
\frac{\epsilon_0\rho({\bf r})}{\alpha^2({\bf r})} \right]  
\frac{z[{\bf u}(k,z^{\ast};{\bf r}')]^{\ast}{\bf u}(k,z;{\bf r})} 
{h(z;{\bf r}')} = 
\frac{\epsilon_0 \rho({\bf r})}{\alpha^2({\bf r})}
\mbox{{\boldmath $\delta$}}({\bf r}'-{\bf r}). 
\label{B5}
\end{equation}
We apply (\ref{B1}) to the first term between square brackets. Subsequently, 
we replace $C_2$ by $C_3$, so that (\ref{23}) comes into play again. 
Eventually, the integrals $\int dz z /[h(z;{\bf r})h(z;{\bf r}')] =0$ and 
$\int dz z/[2\pi i h(z;{\bf r})] =1$ must be proved. As before, this can be 
done by passing over from the contour $C_3$ to the large circle $C_2$.


\section*{References}
\begin{thebibliography}{20}

\bibitem{HB:1992} Huttner B and Barnett S M 1992 
{\em Europhys. Lett.} {\bf 18} 487 \\ 
Huttner B and Barnett S M 1992 {\em Phys. Rev.} A {\bf 46} 4306  

\bibitem{JAU:1948} Jauch J M and Watson K M 1948 {\em Phys. Rev.} 
{\bf 74} 950 \\ 
Jauch J M and Watson K M 1948 {\em Phys. Rev.} {\bf 74} 1485 

\bibitem{FAN:1961} Fano U 1961 {\em Phys. Rev.} {\bf 124} 1866 

\bibitem{KNO:2001} Kn\"{o}ll L, Scheel S and Welsch D G 2001 
{\em Coherence and Statistics of Photons and Atoms}, 
ed J~Pe\v{r}ina (New York: Wiley) p 1 

\bibitem{LUK:2002} Luk\v{s} A and Pe\v{r}inov\'{a} V 2002 
{\em Progress in Optics} vol XLIII, ed E Wolf 
(Amsterdam: North-Holland) p 295  

\bibitem{YG:1996} Yeung M S and Gustafson T K 1996 
{\em Phys. Rev.} A {\bf 54} 5227  

\bibitem{DKW:2002} Dung H T, Kn\"{o}ll L and Welsch D G 2002 
{\em Phys. Rev.} A {\bf 65} 043813 \\
Dung H T, Kn\"{o}ll L and Welsch D G 2002 
{\em Phys. Rev.} A {\bf 66} 063810

\bibitem{SWO:2004} Suttorp L G and van Wonderen A J to be published 

\bibitem{SW:2004} Suttorp L G and Wubs M 2004 {\em Phys. Rev.} A 
to be published 

\bibitem{HEI:1954} Heitler W 1954 {\em The quantum theory of 
radiation} 3rd edn (Oxford: Clarendon) p~38 

\bibitem{POW:1964} Power E A 1964 {\em Introductory Quantum 
Electrodynamics} (London: Longmans) sec. 2.3 

\bibitem{LOU:1973} Louisell W H 1973 {\em Quantum Statistical Properties of 
Radiation} (New York: Wiley) p 240, p 91   

\bibitem{MF:1953} Morse P M and Feshbach H 1953 {\em Methods of Theoretical 
Physics} (New York: McGraw-Hill) p~883 

\bibitem{GOL:1980} Goldstein H 1980 {\em Classical Mechanics} 2nd edn  
(Reading, Mass.: Addison-Wesley) p 567 

\bibitem{SEN:1960} Senitzky I R 1960 
{\em Phys. Rev.} {\bf 119} 670 

\bibitem{FOR:1965} Ford G W, Kac M and Mazur P 1965 
{\em J. Math. Phys.} {\bf 6} 504

\bibitem{ULL:1966} Ullersma P 1966 {\em Physica} {\bf 32} 27 

\bibitem{DEK:1981} Dekker H 1981 
{\em Phys. Rep.} {\bf 80} 1 

\bibitem{JAC:1975} Jackson J D 1975 {\em Classical Electrodynamics} 2nd edn  
(New York: Wiley)  

\bibitem{HAA:1985} Haake F and Reibold R 1985 
{\em Phys. Rev.} A {\bf 32} 2462 

\bibitem{ROM:1997} Romero L D and Paz J P 1997 
{\em Phys. Rev.} A {\bf 55} 4070 

\bibitem{FOR:2001} Ford G W and O'Connell R F 2001 
{\em Phys. Rev.} D {\bf 64} 105020   

\bibitem{BRA:2002} Braun D, Haake F and Strunz W T 2001 
{\em Phys. Rev. Lett.} {\bf 86} 2913 \\
Braun D 2002 {\em Phys. Rev. Lett.} {\bf 89} 277901 \\
Strunz W T and Haake F 2003 {\em Phys. Rev.} A {\bf 67} 022102  

\bibitem{CON:1978} Conway J B 1978 {\em Functions of One Complex Variable} 2nd edn 
(New York: Springer) p 123 

\end{thebibliography}
\end{document}